\documentclass [twocolumn,pra,showpacs,preprintnumbers,amsmath,amssymb,showkeys]{revtex4}

\usepackage{graphicx}
\usepackage{dcolumn}
\usepackage{bm}

\begin{document}

\preprint{APS/123-QED}

\bibliographystyle{apsrev}
\title{An algebraic approach to the Tavis-Cummings problem\\}

\author{I.~P.~Vadeiko}
\email{vadeiko@mkk.ifmo.ru} \affiliation{Department of
Mathematics, Saint-Petersburg Institute of Fine Mechanics and
Optics, 14 Sablinskaya str., St. Petersburg,  197101, Russia}

\author{G.~P.~Miroshnichenko}
\email{mirosh@mkk.ifmo.ru} \affiliation{Department of Mathematics,
Saint-Petersburg Institute of Fine Mechanics and Optics, 14
Sablinskaya str., St. Petersburg,  197101, Russia}

\author{A.~V.~Rybin}
\email{rybin@phys.jyu.fi} \affiliation{Department of Physics,
University of Jyv\"{a}skyl\"{a}, PO BOX 35, FIN--40351,
Jyv\"{a}skyl\"{a}, Finland.}

\author{J.~Timonen}
\email{jussi.timonen@phys.jyu.fi} \affiliation{Department of
Physics, University of Jyv\"{a}skyl\"{a}, PO BOX 35, FIN--40351,
Jyv\"{a}skyl\"{a}, Finland.}

\date{\today}

\begin{abstract}
An algebraic method is introduced for an analytical solution of
the eigenvalue problem of the Tavis-Cummings (TC) Hamiltonian,
based on polynomially deformed \textit{su(2)}, i.e. $su_n(2)$,
algebras. In this method the eigenvalue problem is solved in terms
of a specific perturbation theory, developed here up to third
order. Generalization to the ${\cal N}$-atom case of the Rabi
frequency and dressed states is also provided. A remarkable
enhancement of spontaneous emission of ${\cal N}$ atoms in a
resonator is found to result from collective effects.
\end{abstract}


 \maketitle

\section{Introduction}

We consider here the collective behavior of the system of ${\cal
N}$ two-level atoms coupled to a single mode of the
electromagnetic field in a resonator. The useful form for the
atom-field interaction was proposed in the rotating wave
approximation (RWA) by Tavis and Cummings \cite{Tavis:1968}. In
their model ${\cal N}$ identical two-level atoms interact via
dipole coupling with a single-mode quantized radiation field at
resonance, so that the Hamiltonian is given by

\begin{eqnarray}
&H = H_0 + V,\quad H_0 = \omega \cdot a^\dagger a + \omega
_0 \left( {S_3 + \frac{{\cal N} }{2}} \right),\;\nonumber \\
& V = g\left( {a^\dagger \cdot S_ - + a \cdot S_ + } \right).
\label{eq1}
\end{eqnarray}

\noindent Here  $\omega $ is a frequency of the electromagnetic
field and $\omega _0 $ is the level splitting of the two-level
atoms. The operators $S_3, S_\pm$ are collective spin variables of
${\cal N}$ two-level atoms. These operators are defined as
\begin{eqnarray}
S_3 = \sum\limits_{j=1}^{{\cal N}}\sigma_3^j,\; S_\pm =
\sum\limits_{j=1}^{{\cal N}}\sigma_\pm^j, \label{eq_sigma}
\end{eqnarray}
where $\sigma$'s are Pauli matrices. They satisfy the
\textit{su(2)} algebra. $a, a^\dagger$ are the annihilation and
creation operators of the field. Due to historical reasons  the
Tavis-Cummings (TC) model is often called the Dicke
model~\cite{Dicke:1954}. We concentrate here on the case of exact
resonance, i.e. $\omega = \omega _0 $. In this case the system
exhibits a most interesting collective behavior. For simplicity,
time will be measured in units of the coupling constant $g$, i.e.
we assume in the following that $g = 1$.

The  one-atom version of the TC model, i.e. the Jaynes-Cummings
model(JC)~\cite{Jaynes:1963} has recently attracted much attention
because of e.g. the spectacular experiments reported in
\cite{Meschede:1985}. In this work the properties of a dilute flux
of atoms excited to Rydberg states and interacting with a single
quantum mode of the electromagnetic field were analyzed in a
high-Q cavity. The results reported in \cite{Meschede:1985}
demonstrate that the physical assumptions underlying the JC model
can indeed be experimentally realized. This triggered a lot of
theoretical interest towards the one-atom micromaser(microlaser)
(see \cite{Filipowicz:1996,Elmfors:1996,Raimond:2001} and
references therein). The  analysis of the JC model is greatly
simplified by the fact that  the eigenvalue problem of the
Jaynes-Cummings Hamiltonian can be solved exactly. An algebraic
approach to the solution of generalized Jaynes-Cummings types of
models can be found in~ \cite{Sixia:1997,Yang}. The approach to
the Tavis-Cummings type models based on the Algebraic Bethe Ansatz
is formulated in \cite{Rybin:1998}.

An analytical solution of the ${\cal N}$-atom case, i.e. the TC
model, is much more complicated than that of JC model and is thus
still far from completion. In this article we analyze this model
and develop an analytical technique that is applicable to a
variety of problems of TC type.

To begin with we notice that Hamiltonian (\ref{eq1}) belongs to a
class of operators that can be expressed in the form:
\begin{eqnarray}
H = f(A_0)+g\cdot\left( {A_+ +A_-} \right). \label{H_g}
\end{eqnarray}
Here $f(x)$ is an analytic function of $x$ with real coefficients,
while the operators $A_\pm$ satisfy  the commutation relations
\begin{equation}
\label{eq2} \left[ {A_0 ,A_\pm } \right] = \pm A_\pm.
\end{equation}
We also assume the operator $H$ to be self-adjoint.  This means
that, for any irreducible representation (irrep), the operators
$A_0, A_\pm$ must satisfy the conditions

\begin{equation}
\label{ad_con} \left( {A_0}\right )^\dagger=A_0,
\;\left({A_-}\right )^\dagger=A_+.
\end{equation}
It is worth mentioning that Hamiltonians of type (\ref{eq1}) are
also used in description of Raman and Brillouin scattering, and
frequency conversion, as well as parametric amplification that
involves trilinear boson operators\cite{Carusotto:1989}:

\begin{equation}
{\mathop{\rm H}\nolimits}  = \omega a^\dagger a + \sum\limits_{i =
1}^2 {\omega _i \cdot b_i^ \dagger  b_i }  + g\left( {b_1^ \dagger
b_2 a + b_1 b_2^ \dagger a^ \dagger  } \right)\label{lin3}
\end{equation}
The equivalence of Hamiltonians (\ref{eq1}) and (\ref{lin3}) can
be readily seen by applying the Schwinger transformation
\begin{equation}
S_ +   = b_1^ \dagger  b_2 ,\;S_ -   = b_1 b_2^ \dagger  ,\;S_3  =
{\frac12}\left( {b_1^ \dagger  b_1  - b_2^ \dagger  b_2 } \right).
\label{Schwinger}
\end{equation}

The Hamiltonians of the type of Eq.~(\ref{H_g}) are usually
analyzed by approximating them with an exactly solvable one. The
solvable Hamiltonian is usually quadratic in the boson operators
or linear in the generators of a classical Lie algebra. In this
approach the so-called parametric approximation or its variations
\cite{Smithers:1974, Mollow:1967, Tucker:1969} are often used. The
basic assumption of this approximation is  that one of the quantum
modes or subsystems is prepared in a highly excited (often
coherent) state,  or in a state close to the  vacuum. This
approach, however, puts certain restrictions  on the type of
possible initial conditions as well as on the time span over which
the quantum dynamics can be followed. Another type of an approach
to the solution of Hamiltonian (\ref{H_g}) is based on
perturbation theory for nonlinear algebras \cite{Klimov:2000}.
This approach requires however the existence of a small parameter.
In the TC model such a parameter was found in the case the atoms
are in completely symmetric states ($r=\frac{\cal N}2$)
\cite{Kozierowski:1990,Chumakov:1994}.

The basic idea of the present work is to combine these two
approaches through  algebraic methods as applied e.g. in
\cite{Saavedra:1998,Brief:1996, Sunilkumar:2000, Delgado:2001}).
To this end we reformulate the Hamiltonian in terms of an algebra
that better allows the diagonalization of the Hamiltonian. This
idea was already used by Holstein and Primakoff
\cite{Holstein:1940}. They expressed the generators $S_3, S_\pm$
of the \textit{su(2)} algebra in terms of boson operators $b,
b^\dagger$,
\begin{eqnarray}
S_3  = r - b^ \dagger  b ,\, S_ +   = \sqrt {2r} \sqrt {1 -
\frac{{b^ \dagger b}}{{2r}}}  \, b,\;S_ -   = (S_ + )^\dagger.
\label{H_P}
\end{eqnarray}
Here $r$ is an index that characterizes the irrep of
\textit{su(2)}. However, in \cite{Holstein:1940} the square root
in the transformation Eq. (\ref{H_P}) was in the end replaced by
unity, which amounts to applying the so-called "weak field"
approximation ($\left\langle {b^\dagger b} \right\rangle \ll 2r$).
Obviously, this approximation corresponds to zeroth order in the
expansion of the problem with respect to parameter $\frac 1{2r}$.
Transfromation (\ref{H_P}) has also been applied
\cite{Persico:1975} with expansion up to second order.

We consider here the case when the operators $A_0, A_\pm$ in
Eq.~(\ref{H_g}) are generators of a polynomial deformation
$su_n(2)$ of the Lie algebra \textit{su(2)}
\cite{Daskaloyannis:1993,Sunilkumar:2000,Ruan:2001}.  Numerous
physical applications exist for polynomially deformed algebras
\cite{Karassiov:1994, Holstein:1940, Brandt:1979, Rasetti:1972,
Katriel:1979, Gerry:1971, Katriel:1986, Carusotto:1988,
Lenis:1993, Shanta:1994, Floreanini:1996, Ruan:1999}.  A
particularly interesting, in view of the present problem,
application of deformed algebras was developed by Karassiov (see
in \cite{Karassiov:1994} and refs. therein). The method to be
introduced below is an extension of Karassiov's method.

We introduce here the notion of a Polynomial Algebra of
Excitations (PAE).  In this algebra the coefficients of the
structure polynomials are $c$-numbers, rather than Casimir
operators as is typical of polynomial deformations. We derive an
exact mapping between isomorphic representations of two arbitrary
PAE. We also provide a classification of isomorphic
representations of Polynomial Algebras of Excitations. In our
approach classes of isomorphic representations are specified by
the multiplicity of the maximal and minimal eigenvalues of the
operator $A_0$ in the given representation. We formulate then an
analytical approach that allows us to expand the Hamiltonian, when
expressed in terms of PAE, as a perturbation series.

For completely symmetric states of atoms our results agree with
those reported in  \cite{Kozierowski:1990,Chumakov:1994}. Our
formalism provides however a solution of the problem for any value
of $r$, which allows us to discuss new physical effects in the
Dicke model.

The article is organized as follows. In the next Section II we
discuss the irreducible representations of PAE. In the Section III
we apply the general approach to the Tavis-Cummings model, and in
Section IV construct the perturbation theory for the TC
Hamiltonian and solve its eigenvalue problem up to third order.
The generalized $\cal N$-atom quantum Rabi frequency is defined
for arbitrary quantum states of the system. In the Section V we
use the zero order approximation for the TC Hamiltonian to
calculate the intensity of spontaneous emission of atoms prepared
in the state of thermal equilibrium with the resonator mode. We
show that the correlation of the atoms due to interaction with the
field gives rise to the enhancement of spontaneous emission as
compared to the atoms in the absence of resonator. In conclusion,
we discuss possible further applications of the methods developed
here. Technical details of the algebraic manipulations are given
in the Appendix.

\section{Irreducible representations of the Polynomial Algebra
of Collective Excitations}
\label{sec:part1}

The coefficients of the structure polynomial of a polynomially
deformed algebra are  usually expressed  through the Casimir
operators of the algebra. In this Section we discuss
representations of a special class (PAE) of polynomially deformed
algebras when the coefficients of the structure polynomial are
$c$-numbers. We denote a PAE with a structure polynomial of order
$\kappa$ as $\mathfrak{U}_\kappa$. Formally $\mathfrak{U}_\kappa$
is an associative algebra with unity, defined by three generators
$A_\pm,A_0 $. These generators satisfy two basic commutation
relations, Eq.~(\ref{eq2}). As can be readily seen from these
commutation relations, $\left[ {A_0 ,A_ + A_ - } \right] = 0$. We
can thus assume that
\begin{equation}
\label{eq3} A_ + A_ - = p_\kappa \left( {A_0 } \right)=c_0
\prod\limits_{i = 1}^{\kappa}  {\left( {A_0 - q_i } \right)}.
\end{equation}

\noindent Here $p_\kappa \left( x \right)$ is a structure
polynomial of order $\kappa$, whose coefficients are generally
complex numbers. The terminology is chosen in analogy to the
structure functions of quantum algebras ($q$-deformed algebras)
\cite{Bonatsos:1993}, and the structure constants of the linear
Lie algebras. The set of $\kappa$ real   roots of the structure
polynomial is denoted as $\left\{ {q_i } \right\}_{i=1}^\kappa$.
In physical applications the operators $A_\pm$ of Eq.~(\ref{eq2})
often play the role of creation and annihilation operators of
collective excitations. Therefore, hereafter the algebra
$\mathfrak{U}_\kappa$ will be referred to as the polynomial
algebra of excitations (PAE) of order $\kappa$. Notice that this
algebra is different from the algebra $su_n(2)$. All PAE's are
completely defined by $\kappa+1$ c-numbers, the coefficient $c_0$
and the $\kappa$ roots $\left\{ {q_i } \right\}$ of the structure
polynomial. In the case of  $su_n(2)$, however, the structure
polynomial has some coefficients in the form of Casimir operators.

Below we consider two elementary, but important for the following,
examples of PAE.
 To begin we  assume, without loss of generality,  that $c_0 = \pm 1$.
 Indeed, in the case $\left| {c_0 } \right| \ne 1$, it is always
 possible to renormalize the generators of $\mathfrak{U}_\kappa$,

\begin{equation}
\label{eq4}
A_\pm \to \left| {c_0 } \right|^{ - \frac{1}{2}}A_\pm ,
\end{equation}

\noindent such that the commutation relations Eq.~(\ref{eq2})
remain intact. Using these commutation relations it can also be
readily seen that

\begin{eqnarray}
A_0 A_\pm = A_\pm \left( {A_0 \pm 1} \right),\;  A_- A_ + =
p_\kappa \left( {A_0 + 1} \right). \label{eq5}
\end{eqnarray}

\noindent  As indicated above, $A_ + $ and $A_ - $ have the
physical meaning of creation and  annihilation operators of
collective excitations (quasiparticles), while $A_0 $ is the
operator for the number of excitations. The most simple and
important example of a PAE of first order, $\mathfrak{U}_1$, is
provided by the well-known Heisenberg-Weil Lie algebra, viz.

\begin{eqnarray}
\label{eq6} &b^ \dagger \rightarrow A_ + ,\;b\rightarrow A_- ,\;
b^\dagger b \rightarrow A_0 ,\nonumber\\
& c_0 = 1, q_1=0.
\end{eqnarray}
Here $b, b^\dagger$ are the usual boson operators. For the sake of
simplicity, in what follows we will denote the generators of
$\mathfrak{U}_1$ by $b, b^\dagger$. The algebra $\mathfrak{U}_1$
allows us  to construct the irrep of any other PAE of higher order
$\kappa>1$ as a multiple tensor product of $\mathfrak{U}_1$.

As practical example let us describe in detail a second order PAE
denoted in the paper by $\mathbb{S}_r$. It is relevant to the
algebra \textit{su(2)} given by the commutation relations
\begin{equation}
\label{eq22} \left[ {S_3 ,S_\pm } \right] = \pm S_{\pm ,\quad }
\left[ {S_ + ,S_ - } \right] = 2S_3.
\end{equation}

\noindent It is plain that
\begin{equation}
\label{Cas} S_ + S_ - = S^2 - S_3 ^2 + S_3.
\end{equation}
For every irrep the Casimir operator $S^2$ is equal to $r\left( {r
+ 1} \right)I$, where $I$ is the identity operator. The
corresponding PAE of second order, $\mathbb{S}_r$,  is constructed
such that
\begin{eqnarray}
& S_ +\rightarrow A_ +, S_ -\rightarrow A_ -, S_3\rightarrow
A_ 0 ; \nonumber\\
& c_0 = - 1,\;q_1 = -r,\;q_2 = r+1. \label{eq23}
\end{eqnarray}
Obviously, $\mathbb{S}_r$ has a matrix irrep isomorphic to the
irrep of \textit{su(2)} with the same $r$. Notice that {\it only}
in this representation of $\mathbb{S}_r$ the condition
Eq.(\ref{ad_con}) is fulfilled. This gives us a motivation to
denote the three generators of $\mathbb{S}_r$ as
$\tilde{S}_3,\tilde{S_\pm}$.

Below we describe a general method to construct a realization of
any PAE $\mathfrak{U}_\kappa$ through the algebra
$\mathfrak{U}_1$. Choosing a root $q_j$ of the structure
polynomial we can construct this realization in the form

\begin{eqnarray}
& A_0 = b^\dagger_j b_j + q_j,\; A_ + = \sqrt {c_0 \prod\limits_{i
= 1,i\ne j}^n {\left( {b^\dagger_j b_j + q_j - q_i
} \right)} }\; b_j^\dagger, \nonumber \\
& A_ - = b_j\;\sqrt {c_0 \prod\limits_{i = 1,i \ne j}^n {\left(
{b^\dagger_j b_j + q_j - q_i } \right)} }. \label{eq9}
\end{eqnarray}

\noindent Here $b_j,b_j^\dagger$ are the usual boson operators
associated with the chosen root $q_j$ of the structure polynomial
$p_\kappa \left( x \right)$. The corresponding Fock vectors, i.e.
the eigenvectors of $b^\dagger_j b_j$, are denoted as
$|n\rangle_j$. In what follows the chosen root $q_j$ of the
structure polynomial will be referred to as the {\it pivotal}
root.
 Notice that the product in (\ref{eq9}) always contains exactly
$\kappa$ multipliers regardless of the multiplicity of the root.
Since $b^\dagger_j b_j$ has a well-defined discrete spectrum, the
square root function is defined in the form of spectral
decomposition. We specify the branch of the square root by
choosing $\sqrt { - 1} = i$. It is easy to see that the condition
Eq.(\ref{eq3}) and the commutation relations Eq.(\ref{eq2}) for
generators of $\mathfrak{U}_\kappa$ are satisfied.

A useful automorphism $\hat {T}$,
\begin{equation}
\label{eq11} \hat {T}b^ \dagger = i \cdot b, \; \hat {T}b = i
\cdot b^ \dagger \Rightarrow \hat {T}b^\dagger b = - \left(
{b^\dagger b + 1} \right),
\end{equation}
of $\mathfrak{U}_1$ allows us to construct another realization of
$\mathfrak{U}_\kappa$ through $\mathfrak{U}_1$,
\begin{widetext}
\begin{eqnarray}
A_0 = q_j  - 1 - b^\dagger_j b_j,\quad A_ + = b_j\sqrt {\left( { -
c_0 } \right)\prod\limits_{i = 1,i\ne j}^\kappa {\left( {q_j - q_i
- b^\dagger_j b_j}
 \right)} } ,\quad
A_ - = \sqrt {\left( { - c_0 } \right)\prod\limits_{i = 1,i\ne
j}^\kappa {\left( {q_j - q_i - b^\dagger_j b_j}
 \right)} }\; b_j^ \dagger. \label{eq12}
\end{eqnarray}
\end{widetext}

It is worth mentioning that transformations similar to
Eqs.~(\ref{eq9}), (\ref{eq12}) have been introduced earlier under
the name of multiboson realizations of Bose operators.  These
multiboson realizations satisfy the usual boson commutation
relations, $\left[ {A,A^ \dagger } \right] = 1$
\cite{Rasetti:1972,Brandt:1979,Katriel:1979}.

Applying the realizations Eqs.(\ref{eq9}),(\ref{eq12}) in any
representation of $\mathfrak{U}_1$, we can construct a
representation of $\mathfrak{U}_\kappa$.  In the case of
realization Eq.(\ref{eq9}), an irreducible representation of
$\mathfrak{U}_\kappa$ is constructed through application of the
operator $A_+$ to the vacuum vector $|0\rangle_j$. The finite
dimensional representation can be constructed in the case when the
root $q_{j+1}$ is separated from $q_j$ by a natural number $d$.
Then it can be readily seen that $A_+^d|0\rangle_j=0$. Indeed,
provided that $A_-=A_+^\dagger$, it is not difficult to show that
the norm of the vector $A_+^d|0\rangle_j$ vanishes, i.e.
${}_j\langle 0| A_-^dA_+^d|0\rangle_j=0$. The general construction
can be exemplified by $\mathbb{S}_r$. When $r$ is integer or
half-integer, $d=q_1 - q_2 = 2r + 1$ is integer, and
$\mathbb{S}_r$ has a finite dimensional irrep which is isomorphic
to the corresponding irrep of \textit{su(2)}.

In the case of realization Eq.(\ref{eq12}), the corresponding
irrep is constructed through application of the operator $A_-$ to
the vacuum vector $|0\rangle_j$. Using an argumentation similar to
that given above, it can be seen that the finite dimensional
representation can now be constructed provided that the root
$q_{j-1}$ is separated from $q_j$ by a natural number.

The   meaning of transformations Eqs.(\ref{eq9}) and (\ref{eq12})
becomes now transparent .  The realization Eq.(\ref{eq9})
corresponds to the case when the operator $A_+$ is a creation
operator, while the realization Eq.(\ref{eq12}) corresponds to the
case when $A_+$ is an annihilation operator.

Since the spectrum of operator $b^\dagger_j b_j$ is a set of
natural numbers and zero, the operator $A_0 - q_j $ in Eq.
(\ref{eq9}) has a non-negative spectrum. Therefore, the argument
of the square root function is a positive operator in the finite
dimensional subspace, where the structure polynomial
$p_\kappa(A_0)$ has non-negative spectrum. In this case the
operators $A_ + $ and $A_ - $ are Hermitian conjugated.  In the
subspace corresponding to the negative values of the spectrum of
the structure polynomial, the argument of the square root function
in Eq.(\ref{eq9}) has a negative spectrum. The operators will be
anti-conjugated, i.e. $ \left( {A_ - } \right)^ \dagger = - A_ +$,
which is not plausible. Thus, in a physical problem (see
(\ref{ad_con})), we should only consider those irrep for which the
spectrum of $p_\kappa(A_0)$ is nonnegative.  The same is true for
Eq.~(\ref{eq12}).

The two relations, Eqs.(\ref{eq9}) and (\ref{eq12}), between any
two algebras $\mathfrak{U}_\kappa$ and $\mathfrak{U}_1$, show that
there is no principal difference as to how exactly the meanings of
the creation and annihilation operators of collective excitations
are prescribed to the pair $A_\pm$. The important point is that
the pair exists. It is the physical problem in question which
prescribes the meaning of operators $A_\pm$ and determines the
location of the equidistant spectrum of $A_0 $ on the real axis.
Should one be interested in the eigenvalues of $A_0$, to the right
of the pivotal root $q_j $, it is necessary to choose the
transformation Eq.(\ref{eq9}), while for the region to the left of
$q_j $, it is necessary to use the realization (\ref{eq12}). As
was explained above, we choose the region such that the structure
polynomial is nonnegative.

The general considerations given above can be illustrated by the
$\mathbb{S}_r$ algebra. In the case of the realization of
$\mathbb{S}_r$ connected to the pivotal root $q_1=-r$ it is
necessary to use Eq.(\ref{eq9}). This leads to the conventional
Holstein-Primakoff representation,

\begin{eqnarray}
\tilde{S}_3 = b^\dagger_1 b_1 - r,\, \tilde{S}_ + = \sqrt {2r + 1
- b^\dagger_1 b_1 } \;b_1^ \dagger , \, \tilde{S}_ - =
(\tilde{S}_+)^\dagger . \;\label{eq25}
\end{eqnarray}

\noindent In the case of $q_2=r+1 $, we use the realization
Eq.~(\ref{eq12}),
\begin{eqnarray}
\tilde{S}_3 = r - b^\dagger_2 b_2 ,\, \tilde{S}_ + = b_2 \sqrt {2r
+ 1 - b^\dagger_2 b_2 },\, \tilde{S}_ - = (\tilde{S}_+)^\dagger
.\; \label{eq24}
\end{eqnarray}

\noindent The spectra of $b^\dagger_1 b_1 $ and $b^\dagger_2 b_2 $
are limited from above by the value $2r$, while the subspace
spanned by eigenvectors $\left| n \right\rangle_{1,2} \;\left( {n
= 0,1,\ldots 2r} \right)$ of operator $A_0 $ is irreducible (cf.
Eq.~(\ref{H_P})).

Obviously all the realizations of PAE constructed through
Eqs.~(\ref{eq9}), (\ref{eq12}) are fully  characterized by the
dimension $d$ of the invariant subspace  and by the order $k_-$ of
the left and $k_+$ of the right roots defining the corresponding
irreducible representation. Thus an irrep of PAE is characterized
by a set of parameters $\left\{k_-,k_+,d \right\}$. Such irrep we
will denote by $R(k_-,k_+,d)$. For instance, $R(1,0,\infty)$ means
the representation of $\mathfrak{U}_1$, while the irrep of
$\mathbb{S}_r$ is $R(1,1,2r+1)$.

An isomorphism between irreps of $\mathfrak{U}_\kappa$ and
$\mathfrak{U'}_{\kappa'}$ that belong to the same class
$R(k_-,k_+,d)$, is given by
\begin{eqnarray}
&A_0 = A'_0 + (q_j - q'_{j'}) ,\nonumber\\
& A_+ = \sqrt{\frac{c_0 \prod\limits_{i = 1}^\kappa {\left( {A'_0
+q_j -q'_{j'} - q_i } \right)} } {c'_0 \prod\limits_{i' =
1}^{\kappa'}
{\left( {A'_0 - q'_{i'} } \right)} }}A'_+ ,\nonumber \\
& A_- = A'_ - \sqrt{\frac{c_0 \prod\limits_{i = 1}^\kappa {\left(
{A'_0 +q_j -q'_{j'} - q_i } \right)} } {c'_0 \prod\limits_{i' =
1}^{\kappa'} {\left( {A'_0 - q'_{i'} } \right)} }}, \label{eq13}
\end{eqnarray}

\noindent or by
\begin{eqnarray}
&A_0 = (q'_{j'} + q_j-1)-A'_0 , \nonumber\\
&A_+ = A'_- \sqrt{\frac{c_0 \prod\limits_{i = 1}^\kappa {\left(
{q'_{j'} +q_j -A'_0 - q_i } \right)} } {c'_0 \prod\limits_{i' =
1}^{\kappa'}
{\left( {A'_0  - q'_{i'} } \right)} }},\nonumber \\
& A_- = \sqrt{\frac{c_0 \prod\limits_{i = 1}^\kappa {\left(
{q'_{j'}+q_j -A'_0 - q_i } \right)} } {c'_0 \prod\limits_{i' =
1}^{\kappa'} {\left( {A'_0  - q'_{i'} } \right)} }}\;A'_ + .
\label{eq14}
\end{eqnarray}
In Eqs.~(\ref{eq13}),(\ref{eq14}) the operator argument of the
square root function should be taken after identical multipliers
in the nominator and denominator are cancelled. The pivotal roots
$q_j$ and $q'_{j'}$ define a vacuum vector of the irrep. Compare
these expressions with Eqs.(\ref{eq9}),(\ref{eq12}).

To recapitulate: in the general case two irreducible
representations $R(k_-,k_+,d)$ and $R'(k'_-,k'_+,d')$ of
$\mathfrak{U}_\kappa$ and $\mathfrak{U'}_{\kappa'}$ are isomorphic
provided that $d=d'$, and
\begin{equation}
k_-=k'_-, k_+=k'_+ \mbox{ or }k_-=k'_+,
k_+=k'_-.\label{eq15}\end{equation}

For symmetric irreps, i.e. when $k_+=k_-$, these conditions
coincide. The isomorphism is a consequence of the fact that, under
conditions (\ref{eq15}), one can choose $q_j$($q'_j$) in such a
way that the  functions under the square root in Eqs.(\ref{eq13}),
(\ref{eq14}) do  not have zeros in the spectrum of operator $A_0$,
and therefore we can consider the square root as a single-valued
analytic function. This means that transformations are invertible,
analytic, and therefore define an isomorphism of irreps.

The constructed transformations give us a tool to realize
$\mathfrak{U}_\kappa$ in terms of a simpler PAE with the same type
of irrep. This procedure will  be applied below to the
Tavis-Cummings Hamiltonian.

\section{The Tavis-Cummings Hamiltonian in terms of
third-order PAE} \label{sec:part2}

 The interaction part of the Hamiltonian Eq.~(\ref{eq1})
can be expressed in terms of third-order PAE. The generators $M_0
,M_\pm $ of this algebra are realized as
\begin{equation}
\label{eq16} M_ - = a \cdot S_ + ,\;M_ + = a^\dagger \cdot S_ -
,\;M_0 = \frac{a^\dagger a - S_3 }{2}.
\end{equation}

\noindent It is plain that these generators satisfy the
commutation relations Eqs.~(\ref{eq2}). The generators of the
algebra $M_0,M_\pm$ commute with the operators

\begin{equation} \label{eq17}
M = a^\dagger a + S_3 + r, \; S^2 = S_3 ^2 +
\frac{1}{2}\left( {S_ + S_ - + S_ - S_ + } \right).
\end{equation}
Hereafter we use the same notation $M$ both for the Casimir
operator and its eigenvalue, if no confusion arises. We show below
that the eigenvalues $M, r(r+1)$ of the operators of
Eq.(\ref{eq17}) parameterize the PAE in question. We thus denote
this PAE as $\mathbb{M}_{M,r}$. The structure polynomial of
$\mathbb{M}_{M,r}$ can be expressed in the form

\begin{eqnarray}
& p_3 \left( {M_0 } \right)=M_ + M_ - = a^\dagger a \left( {S^2 -
S_3 ^2 - S_3 } \right) =  \nonumber
\\ & a^\dagger a \left( {r - S_3 } \right) \left( {r + S_3 + 1} \right) = \nonumber\\
&-\left( {M_0 + \frac{M - r}{2}} \right)\left( {M_0 - \frac{M -
3r}{2}} \right)\left( {M_0 - \frac{M + r+2}{2}} \right).
 \label{eq19}
\end{eqnarray}

\begin{figure}
\includegraphics[width=80mm]{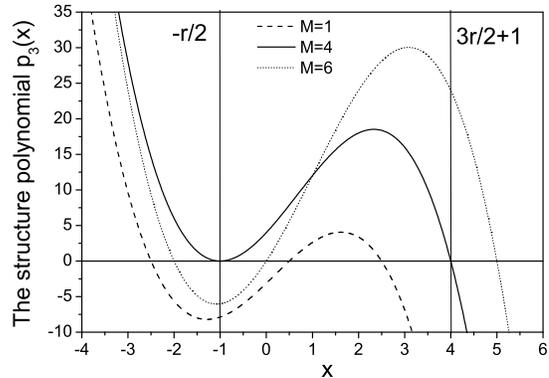}
\caption{\label{fig:polyn3} The structure polynomial $p_3(x)$ of
the $\mathbb{M}_{M,r}$ algebra for $r=2$.}
\end{figure}

\noindent The parameters of this structure polynomial are
\begin{eqnarray}
\label{eq20} &c_0 = - 1,\;q_1 = -\frac{M -r}{2},\nonumber \\
&q_2 =\frac{M - r}{2} -r,\;q_3 = \frac{M - r}{2} + r + 1,
\end{eqnarray}

\noindent and its behavior as a function of $M_0$ is given in
Fig.~\ref{fig:polyn3}.

We turn next  to the description of finite dimensional irrep of
$\mathbb{M}_{M,r}$. In physical applications the parameter $r$ has
the meaning of  collective Dicke index. This index runs from
$\varepsilon({\cal N}) = \frac{1 - \left( { - 1} \right)^{{\cal N}
}}{4}$ to $\frac{{\cal N} }{2}$ with unit steps, while $M$ can be
any natural number including zero. Thus $q_3$ is the biggest
positive root of first order. If $M < 2r$, then $q_1
> q_2 $;  if $M > 2r$ then $q_1 < q_2 $; the case $M = 2r\;
\Rightarrow q_1 = q_2 $  corresponds to a root of second order. A
typical plot of the structure polynomial is shown in
Fig.~\ref{fig:polyn3} for these three cases. A finite dimensional
representation of $\mathbb{M}_{M,r}$ (where $(M_+)^\dagger = M_-$)
corresponds to the positive spectrum of $p_3\left( M_0 \right)$.
The spectrum is limited from the right by $q_3$ and from the left
by $q_1$ or $q_2$. Notice that the number $M$ is conserved.
Therefore this number is determined by the initial state. The
different values of $M$ and $r$ define different algebras
$\mathbb{M}_{M,r}$, whose single physical finite dimensional
representation we will call a {\it zone}. The case $M < 2r$
corresponds to {\it nearby} zones. The two largest roots are $q_1
$ and $q_3 $ and the irrep has the type $R(1,1,M+1)$.
Consequently, the well-known weak-field limit corresponds to
nearby zones.

The case $M> 2r$ corresponds to {\it remote} zones. The two
largest roots are $q_2 $ and $q_3 $, and the corresponding irrep
is of the type $R(1,1,2r+1)$. Notice that the region $M>>2r$ is
usually called the strong-field limit.

In the special case $2r=M$, referred to {\it intermediate} zone,
the algebra $\mathbb{M}_{M,r}$ possesses an irrep of the type
$R(2,1,2r+1)$. It is the only irreducible representation that
principally differs from all the others.

\begin{figure}
\includegraphics[width=85mm]{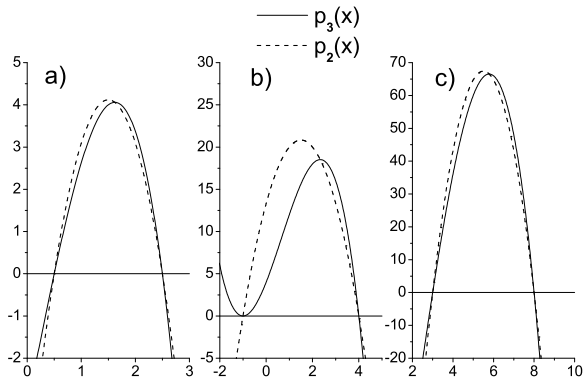}
\caption{\label{fig:polyn23} Approximation of the structure
polynomial $p_3(x)$ by the parabolic $p_2(x)$ for $r=2$. a) $M=1$;
b) $M=4$; c) $M=12$.}
\end{figure}

As indicated above, the simplest PAE with irrep of the type
$R(1,1,d)$ is $\mathbb{S}_{\tilde r}$ (we use here $\tilde{r}$ to
distinguish it from the (physical) collective index $r$). It would
be convenient to solve the eigenvalue problem in terms of the
simplest algebra $\mathbb{S}_{\tilde r}$. Notice that in any
finite dimensional irrep of PAE characterized by $R(1,1,d)$, the
structure polynomial of the algebra can be approximated by a
parabolic curve. This is shown in Fig.~\ref{fig:polyn23} (for $M
\ne 2r$) for the structure polynomial of $\mathbb{M}_{M,r}$. The
larger(smaller) $M$ is in comparison with $2r$, the better is the
approximation. However, for $M \approx 2r$ the approximation is
not satisfactory (see Fig.~\ref{fig:polyn23}b). In the regions
where the approximation is adequate, it is then not difficult to
diagonalize the operator $V=S_ + + S_ - $, defined in terms of
generators of the conventional \textit{su(2)} algebra. The latter
has a parabolic structure polynomial.  In Fig.~\ref{fig:polyn23}
the two roots of $p_2(x)$ are chosen to be  equal to the
corresponding two roots of $p_3(x)$. The choice of $c_0$ in
$p_2(x)$ will be explained below. Thus in the cases of nearby or
remote zones, it is convenient to study the problem in terms of
algebra $\mathbb{S}_{\tilde r}$. The approximation illustrated in
Fig.~\ref{fig:polyn23} indicates that the TC problem can be solved
via an appropriate perturbation theory.

To begin with we consider the  transformation of
$\mathbb{M}_{M,r}$ to $\mathbb{S}_{\tilde r}$ for the case of
remote zones. The dimension of a remote zone is $2r + 1$, and the
algebra $\mathbb{S}_{\tilde r}$ should be characterized by
$\tilde{r}=r$. The finite dimensional irrep of $\mathbb{S}_{\tilde
r}$ is isomorphic to the corresponding irreducible representation
of the atomic subsystem. For the pivotal root $q_j$ we choose the
largest root that bounds the irrep of $\mathbb{M}_{M,r}$ from the
right (the $q_3$ in (\ref{eq20})), while as the root $q'_{j'}$ we
take the $q_1$ of $\mathbb{S}_r$ from Eq.~(\ref{eq23}). Applying
the mapping (\ref{eq14}) we obtain
\begin{eqnarray}
 & M_0 = \frac{M - r}{2} - \tilde {S}_3 ,\;M_ + =\tilde{S}_ - \;  \sqrt
{\left( {M - r +1 - \tilde {S}_3 } \right)}, \nonumber \\
& M_ - = \sqrt {\left( {M - r +1 - \tilde {S}_3 } \right)}\;\tilde
{S}_ + . \label{eq28}
\end{eqnarray}

\noindent The spectrum $\{\tilde {m}\}$ of the operator $\tilde
{S}_3 $ belongs to the region $ - r \le \tilde {m} \le r$,
consequently the argument of the square root function in
Eq.(\ref{eq28}) is positive in the remote zones $(M - r > r)$. The
relations Eqs.(\ref{eq28}) express the generators of algebra
$\mathbb{M}_{M,r}$ as analytic function of the generators of the
$\mathbb{S}_r$ algebra. they thus allow us to approximate the more
complex algebra $\mathbb{M}_{M,r}$ of third order by a simpler
algebra of second order.

Before we begin to study this approximation we construct a
realization of $\mathbb{S}_r$  in terms of boson and spin
variables.
>From Eqs.(\ref{eq16}),(\ref{eq17}), and (\ref{eq28})  it follows
that in remote zones
\begin{equation}
\label{eq29a} \tilde {S}_3 = S_3,\; \tilde {S}_+ =
\frac1{\sqrt{a^\dagger a+1}}\; a\cdot S_+, \quad \tilde {S}_- =
\left ({\tilde {S}_+}\right )^\dagger .
\end{equation}
Notice that the subspaces that correspond to remote zones do not
contain the vacuum state of the field. It is also worth mentioning
that the matrix representation of operator $\frac1{\sqrt{a^\dagger
a+1}}a$ is $\delta_{n,n+1}$ in any remote zone. This operator has
been considered before as a earlier in terms of phase operator
\cite{Chumakov:1994, Saavedra:1998, Delgado:2001}.

We turn now to the nearby zones $M < 2r$. For this region the
mapping of the algebra $\mathbb{M}_{M,r}$ to the algebra
$\mathbb{S}_{\tilde r}$  is realized through procedure similar to
that described above for remote zones. Notice that the dimension
of nearby zones is $d=q_3-q_1=M+1$, and therefore
$\tilde{r}=\frac{M}{2}$ (cf. Fig.~\ref{fig:polyn23}a). For the
nearby zones there is no simple correspondence to a physical
representation like that of Eq.(\ref{eq29a}). The particular
choice of the pivotal root $q'_{j'}$ is unimportant and we use the
same choice as before. Applying Eq.(\ref{eq14}) we obtain
\begin{eqnarray}
& M_0 = \frac{r}{2} - \tilde{S} _3, \,M_ + = \tilde{S} _ - \,\sqrt
{\left( {\frac{4r
- M}{2} +1 - \tilde{S} _3} \right)},\nonumber\\
& M_ + = (M_-)^\dagger . \label{eq34}
\end{eqnarray}
 Since all the eigenvalues of the operator $\tilde{S} _3 $ belong
 to the interval
 $ - \tilde{r}$ to $\tilde{r}$, the argument of the square root function
does not have zero eigenvalues in the nearby zones. The
realization of $\mathbb{S}_{\tilde r}$ through spin and boson
variables is then given by
\begin{eqnarray}
\tilde{S} _3 = \frac M2 - a^\dagger a,\; \tilde{S} _ +
=\frac1{\sqrt{r+1-S_3}} S_+ \, a,\nonumber \\
 \tilde{S} _ -
=S_-\,\frac1{\sqrt{r+1-S_3}} \,a^\dagger. \label{eq34a}
\end{eqnarray}
\noindent Notice that the nearby zones do not contain the
eigenvector $|r,r\rangle$  of $S_3$.

To clarify the structure of intermediate zone we choose as the
pivotal root as the third root of $p_3(x)$ of Eq.(\ref{eq20}), and
apply the transformation (\ref{eq12}). We thus obtain
\begin{eqnarray}
& M_0 = q_3-b^\dagger_3 b_3 -1=\frac{M+r}{2}-b^\dagger_3 b_3, \nonumber \\
& M_ + = b_3 \;\sqrt {\left( {M+1-b^\dagger_3 b_3} \right)
\left( {2r + 1 - b^\dagger_3 b_3 } \right)}\nonumber \\
& M_ - = \sqrt {\left( {M+1-b^\dagger_3 b_3} \right)\left( {2r + 1
- b^\dagger_3 b_3 } \right)} \;b_3^ \dagger . \label{eq_r3}
\end{eqnarray}
Here $b_3, b_3^\dagger$  are generators of the algebra
$\mathfrak{U}_1$.

In the region $M>2r$ the  multiplier $(2r+1-b^\dagger_3 b_3)$ in
Eq.(\ref{eq_r3}) vanishes first. This corresponds to remote zones.
In the region  $M<2r$ the multiplier that vanishes first is
$(M+1-b^\dagger_3 b_3)$. In the intermediate zone $M=2r$ we have
\begin{eqnarray}
M_ + = b_3 \;(2r+1-b^\dagger_3 b_3),\; M_ - = (2r+1-b^\dagger_3
b_3)\;b_3^ \dagger .\, \label{eq_boundary}
\end{eqnarray}
In the intermediate zone we thus obtain a special realization of
the TC Hamiltonian,
\begin{eqnarray}
&H_{\{M=2r\}}=
(2r+\frac12)(b_3+b_3^\dagger)- \nonumber\\
&\frac12 \left[{b^\dagger_3 b_3 (b_3+b_3^\dagger)
+(b_3+b_3^\dagger) b^\dagger_3 b_3}\right]. \label{eq_BModel}
\end{eqnarray}
The domain of the quantum space for this Hamiltonian is specified
by the condition $n_3\le2r$.

\section{Diagonalization of the Tavis-Cummings Hamiltonian}
\label{sec:part4}

Let us introduce an $ \cal N$-atom generalization for arbitrary
values $r$ and $M$ of the well-known quantum Rabi frequency such
that
\begin{equation}
\label{eq_Rabi} \Omega _R \equiv \left\{ {{\begin{array}{c}
  2\sqrt {M - r + \frac{1}{2}}, \;M\geq 2r \\ \\
  2\sqrt {\frac{{4r - M + 1}}{2}}, \; M<2r \\
\end{array}}}\right. .
\end{equation}
For $r=\frac{\cal N}2$ our definition agrees with that used
in\cite{Kozierowski:1990, Chumakov:1994}. Introducing  a small
parameter $\alpha \equiv \left ({\frac 12 \Omega _R}\right )^{-2}$
we can rewrite the realizations of $M_\pm$ in nearby and remote
zones (see Eqs.(\ref{eq28}),(\ref{eq34})) in the form
\begin{equation}
\label{eq29_new} M_+  = \frac{\Omega _R }{2} \tilde S_ -\sqrt {1 -
\alpha\left({\tilde S_3  - \frac 12}\right )}, \; M_- =
M_+^\dagger .
\end{equation}
The diagonalization problem for the operator Eq.(\ref{eq1}) can
now be solved in each zone by means of perturbation theory with
respect to the small parameter $\alpha$. One can show that the
eigenvalues of the argument of the square root function in
Eq.(\ref{eq29_new}) are less then unity. Hence we can expand the
square with respect to $\alpha$, and find thereby for the
interaction part of the Hamiltonian
\begin{widetext}
\begin{eqnarray}
\mathop{\rm V}  = \frac{{\Omega _R }}{2}\left[
\begin{array}{l}
 \left( {\tilde S_ +   + \tilde S_ -  } \right) - \frac{{\alpha }}
 {2}\left( {\left( {\tilde S_3  - \frac12} \right)
 \tilde S_ +   + \tilde S_ -  \left( {\tilde S_3 - \frac12}
 \right)} \right) - \frac{1}{{2!}}\left( {\frac{{\alpha }}{2}} \right)^2
  \left( {\left( {\tilde S_3  - \frac12} \right)^2 \,
  \tilde S_ +   + \tilde S_ -  \,\left( {\tilde S_3  - \frac12}
   \right)^2 } \right) -  \\
 \left( {\sum\limits_{n = 3}^\infty  {\frac{{\left( {2n - 3} \right)!!}}
 {{n!}}\left( {\frac{{\alpha  \left( {\tilde S_3  - \frac12}
  \right)}}{2}} \right)^n } \;\tilde S_ +
   + \tilde S_ -  \;\sum\limits_{n = 3}^\infty  {\frac{{\left( {2n - 3} \right)!!}}
   {{n!}}\left( {\frac{{\alpha  \left( {\tilde S_3  - \frac12}
    \right)}}{2}} \right)^n } } \right) \\
 \end{array} \right]
.\quad \label{eq31}
\end{eqnarray}
\end{widetext}
In the interaction representation the Hamiltonian coincides with
$V$, we only need to diagonalize the latter. Up to third order in
$\alpha$ we find that $$V=\frac{\Omega_R }{2}\left( {{\mathop{\rm
V}\nolimits} ^{\left( 0 \right)}  + {\mathop{\rm V}\nolimits}
^{\left( 1 \right)}  + {\mathop{\rm V}\nolimits} ^{\left( 2
\right)}  + {\mathop{\rm V}\nolimits} ^{\left( 3 \right)} }
\right),$$

\noindent where the $V^{\left( n \right)}$ are terms of
 $n$-th order in $\alpha$, and are given in the
Appendix \ref{app:SimTran}. In the Appendix we also show that
unitary transformations $\tilde U_k ,\quad k = 0,1,2,3$, which
bring the interaction operator into diagonal form:

\begin{widetext}
\begin{eqnarray}
\bar V  \equiv \tilde U \cdot V  \cdot \tilde U^{ - 1} = \Omega_R
\cdot \tilde S_3 \left\{ {1 + \left( {\frac{\alpha }{4}} \right)^2
\cdot \left[ {5\tilde S_3 ^2  - 3\tilde r\left( {\tilde r + 1}
\right) + 1} \right]} \right\}, \quad \tilde U \equiv \tilde U_3
\, \tilde U_2 \,  \tilde U_1 \, \tilde U_0 . \label{eq36}
\end{eqnarray}
\end{widetext}
The spectrum of the operator $V$ as given by Eq.(\ref{eq36})
agrees with the results of \cite{Kozierowski:1990, Chumakov:1994}
for the symmetric states of the atoms.

\begin{figure}
\includegraphics[width=80mm]{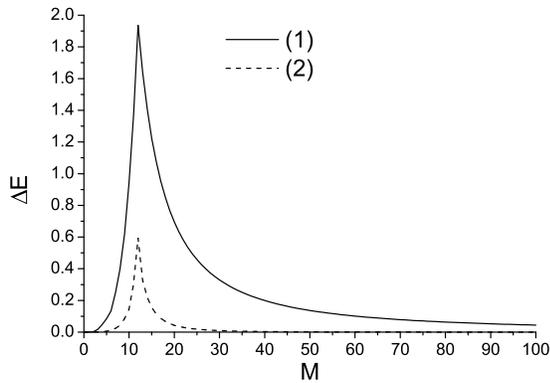}
\caption{\label{fig:spec0} Deviation of the eigenvalues of $V $
from their numerical values in zero (solid line) and second
(dashed line) order in $\alpha$, for $r = 6$.}
\end{figure}

\begin{figure}
\includegraphics[width=80mm]{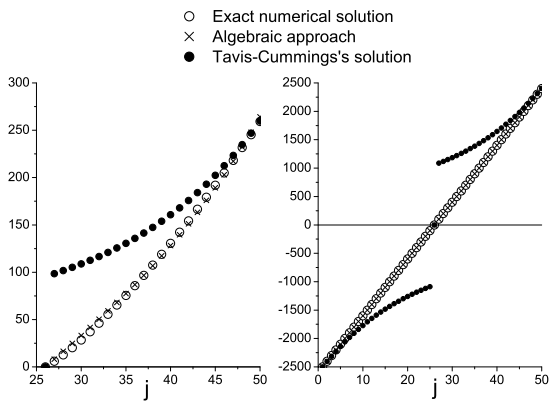}
\caption{\label{fig:spec2}  Energy levels of $V $ in the ascending
order for $r=25$, as calculated numerically (circles), from Eq.
(\ref{eq36}) (crosses), and as is given in \cite{Tavis:1968}
(squares). a) $M=50$; b) $M=2525$. }
\end{figure}

We compared the third order solution Eq.(\ref{eq36}) with the
exact numerical diagonalization of $V$ and found that the result
(\ref{eq36}) is very accurate, especially for increasing values of
$|M-2r|$. The results of this comparison are shown in
Fig.~\ref{fig:spec0}.

In their original article \cite{Tavis:1968} Tavis and Cummings
also found an approximative analytical expression for the spectrum
of the interaction operator $V$. We compare the results of
Ref.~\cite{Tavis:1968} with our analytical and numerical solutions
in Fig.~\ref{fig:spec2}. It is evident that the Tavis and Cummings
solution is only accurate for very large and very small values of
index $j$.

\begin{figure}
\includegraphics[width=80mm]{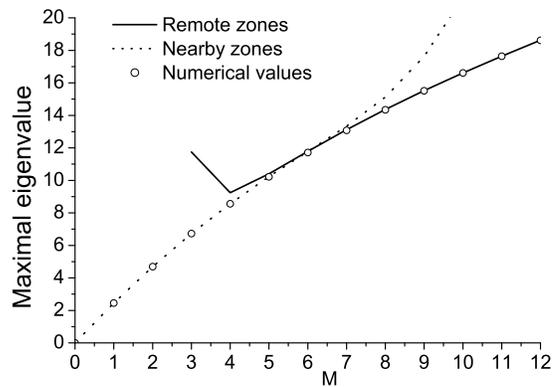}
\caption{\label{fig:spec4} The dependence on $M$ of the maximal
eigenvalue of $V$ for $r = 3$, as calculated numerically and from
Eq.(\ref{eq36}).}
\end{figure}

The Fig.~\ref{fig:spec4} compares the energies calculated
numerically and in accordance with the analytical solution Eq.
(\ref{eq36}). In the intermediate region of $M$ the curves for
nearby and remote zones overlap and coincide thus providing still
satisfactory correspondence to the exact solution. However
evidently the expansion for the remote zone breaks down in the
nearby zone and vice versa. This means that the classification of
zones introduced in this paper is indeed adequate.

\begin{figure}
\includegraphics[width=80mm]{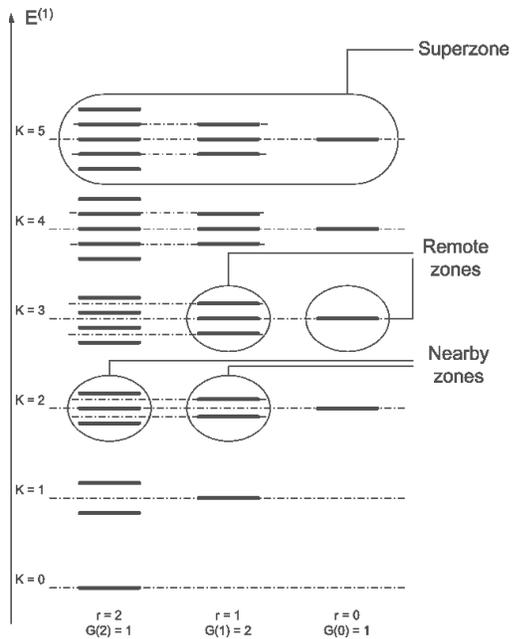}
\caption{\label{fig:sket} Zones structure for ${\cal N}=4$.
$E^{(1)}$ is the spectrum of the zero order Hamiltonian. $K$ is an
eigenvalue of the corresponding operator (\ref{K}), and $G(r)$ is
a number of equivalent irreps of \textit{su(2)} defined below.}
\end{figure}

Finally we consider the ${\cal N}$-atom {\it dressed states}. In
other words we introduce  a representation in which the zero order
Hamiltonian is diagonal. This representation is given by
transformation $\tilde{U}_0$. For ${\cal N}=1$ all the higher
order terms in Eq.(\ref{eq31}) (higher than zero order) vanish,
while the eigenstates of the zero order Hamiltonian coincide with
the dressed states\cite{Scully:1998} of the Jaynes-Cummings model.
We can thus call the eigenstates of the zero order Hamiltonian the
{\it dressed states} of the ${\cal N}$-atom model. Notice also
that in the remote zones the Rabi frequency Eq.(\ref{eq_Rabi})
does not depend on the Dicke index $r$. If we consider only the
zero order terms of $H$, it is convenient to combine all the
remote zones that have the same Rabi frequency into a remote {\it
superzone}, whose dimension is $2^{{\cal N}}$. Introducing an
operator for the total number of quanta in the atom-field system
(cf. the definition of $M$ in Eq.(\ref{eq17})),
\begin{equation}
\label{K} K = a^\dagger a + S_3 + \frac {{\cal N}}2,
\end{equation}
we can define the remote superzone as follows. A remote superzone
contains all eigenvectors of $K$ that have the same eigenvalue,
provided it is  larger than ${\cal N}$. In Fig.~\ref{fig:sket} we
show the definition of the zones in the Tavis-Cummings model for
${\cal N}=4$.

It can be readily seen that, in the Hilbert subspace corresponding
to the remote superzone, the unitary transformation $\tilde{U}_0$
can be factorized into a product of ${\cal N}$ single-particle
transformations that  are the dressing operators for the
Jaynes-Cummings problem (cf. Eq.(\ref{eq29a})),
\begin{equation}
\label{factor} \tilde{U}_0 = \prod\limits_{j=1}^{{\cal N}} {\exp
\left[ {\frac{\pi }{4}\left( {\frac1{\sqrt{a^\dagger a+1}}\; a
\sigma^j_+ - \sigma^j_-a^\dagger\;\frac1{\sqrt{a^\dagger a+1}}}
\right)} \right]}.
\end{equation}

>From the discussion following Eq.(\ref{eq29a}), it follows that in
each remote superzone the field operators $\frac1{\sqrt{a^\dagger
a+1}}\; a$ and $a^\dagger\;\frac1{\sqrt{a^\dagger a+1}}$ commute.
This means that the system behaves almost like a semiclassical
one.

\section{Enhancement of spontaneous emission
in the resonator due to collective effects} \label{sec:part5}

In the previous section we developed an algebraic  approach to the
Tavis-Cummings model. We introduced the  operators $\tilde S_\pm$
describing collective excitations in the atom-field system. In
terms of these operators   we   constructed a perturbation series
for the Tavis-Cummings  Hamiltonian Eq.(\ref{eq1}). The derived
perturbation series gives us a tool to distinguish and classify
cooperative (multiparticle) effects of different orders that are
involved in calculations of different physical observables
characterizing the atom-field system. In the preceding  section we
constructed a Hamiltonian Eq.(\ref{eq31}) corresponding to the
zero order approximation for the Hamiltonian Eq.(\ref{eq1}).  This
simplified operator depends on multiparticle Rabi frequency
Eq.(\ref{eq_Rabi}), which depends on the number of atoms in the
cavity. Therefore the simplified Hamiltonian  Eq.(\ref{eq31})
allows to account for cooperative effects in the system.

In this section we study a contribution of cooperative effects
into the rate of spontaneous emission generated by the atom-field
system.  The atom-field system is assumed to be prepared in the
state of thermal equilibrium.  This state is  described by the
canonical Gibbs ensemble with the thermostat temperature $T$. This
means that under   the "system" we imply $\cal N$ atoms strongly
coupled to the  resonator mode.  Under  the  "thermostat" we imply
the surrounding environment, for instance, cavity walls taken at
the temperature $T$. In the state of thermal equilibrium  the
exact atom-field density matrix should be defined using the
Hamiltonian Eq.(\ref{eq1}). However, for the exact density matrix
the analytical analysis of the transition probability, if possible
at all, would be highly technically involved.

We demonstrate that nontrivial physical results for the intensity
of spontaneous emission of ${\cal N}$ two-level atoms placed
inside the cavity can be already  obtained for the zero order
approximation of the exact Hamiltonian Eq.(\ref{eq31}). The
thermal state is given by
\begin{equation}
\label{Rho_th} \rho_{th}  = \frac1Z\exp\left[ { -\frac{H_0 +
\frac{\Omega_R }{2} \left( {\tilde S_ +   + \tilde S_ - }
\right)}{kT}} \right].
\end{equation}
Here $Z$ is a normalization factor.

We show here that the intensity of spontaneous emission of the
system comprised of $\cal N$ atoms {\it strongly} coupled to the
resonator mode and prepared in  the thermal state, can be greatly
enhanced at a certain temperature. This amplification results from
high correlations in the atomic subsystem.   Similar effect exists
for Dicke's superradiant state \cite{Dicke:1954}. This state is
prepared by a short laser pulse. Therefore the setting of the
Dicke's theory is quite different from our considerations of the
stationary state of thermal equilibrium.

When calculating  the rate of spontaneous emission (or  the
intensity proportional to this quantity) we merely follow the
ideas of Dicke's paper (see e.g. \cite{Dicke:1954}). According to
this theory, the rate of spontaneous emission in the system is
proportional to the average  of the square  of the atomic dipole,
viz.

\begin{equation}
\label{I_spon} I  = I_0 \cdot \langle {S_ + S_ - } \rangle=I_0
\cdot Tr\left\{ {\rho_{th}\cdot S_ + S_ - } \right \}.
\end{equation}
It is  convenient to calculate the average in the dressed states
basis, where
\begin{equation}
\label{Rho_th_dr} \rho_{th} \rightarrow \frac1Z \tilde{U}_0
\rho_{th} \tilde{U}_0^{-1}= \exp\left[ { -\frac{H_0 +
\Omega_R\tilde S_ 3}{kT}} \right].
\end{equation}
Taking into account Eq.(\ref{Cas}) along with the fact that, for
an arbitrary zone,
\begin{equation}
\label{S3} S_3  = (\tilde r -r) + \tilde S_3,
\end{equation}
it can be shown that in the dressed states basis
\begin{eqnarray}
\label{Dipole} &\tilde{U}_0(S_ + S_ -)\tilde{U}_0^{-1}= \tilde r
\left( { 2r-\frac32\tilde r +\frac12 } \right ) - \nonumber
\\ &(r+\frac12- \tilde r)(\tilde{S}_ + \tilde{S}_ -)- \frac14
\left( {\tilde{S}_ +^2 - 2\tilde{S}_3^2 + \tilde{S}_ -^2}\right).
\end{eqnarray}
Finally,  the intensity of spontaneous emission is given by
\begin{eqnarray}
\label{I_spon2} &I  = \frac{I_0}Z \sum\limits_{M=0}^\infty
\sum\limits_{r=\varepsilon}^{\frac{{\cal N}}2}G(r)\nonumber
\\& \sum\limits
_{\tilde m= -\tilde r}^{\tilde r} e^{-\frac{\omega \left( {M-r+
\frac{{\cal N}}2}\right) + \Omega_R\tilde m}{kT}}  \left[{\tilde
r\left( { 2r-\frac32\tilde r +\frac12 } \right ) + \frac12
\tilde{m}^2 }\right].\quad
\end{eqnarray}
Here $\varepsilon\equiv\frac{1-(-1)^{{\cal N}}}4$, and
$G(r)=\frac{{\cal N}! (2r+1)}{(\frac{{\cal N}}2+r+1)! (\frac{{\cal
N}} 2 -r)!}$ is the number of equivalent representations with the
same $r$.

\begin{figure}
\includegraphics[width=80mm]{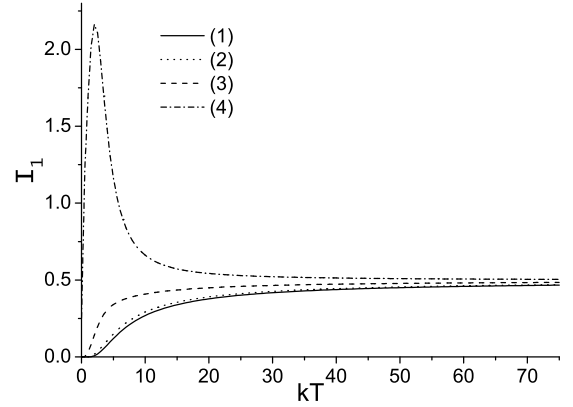}
\caption{\label{fig:Int} The intensity of spontaneous emission per
atom (in units of $I_0$) versus cavity temperature. The curve (1)
is the classical result given by Eq.(\ref{Isl}). The curves
(2,3,4) correspond to ${\cal N}=10$, ${\cal N}=50$, ${\cal
N}=100$, respectively,  and $\frac\omega{g}=10$.}
\end{figure}

Let us consider the intensity per atom, i.e. $I_1\equiv I/{\cal
N}$. This intensity consists of two terms, i.e. the first is given
by a single-particle contribution $I_{single}$ and the second one
proportional to the two-particle correlation function $Cor$,
$I_1=I_{single} + I_0({\cal N}-1)Cor$. They are found to be
\begin{eqnarray}\label{spont1}
&I_{single}\equiv I_0\frac1{\cal N} \langle{
\sum\limits_{i}\sigma_+^i \sigma_-^i} \rangle= I_0 \left(
{\frac12 + \frac1{{\cal N}}\langle{S_3}\rangle}\right) ,\nonumber \\
&Cor\equiv \frac{1}{{{\cal N} \left( {{\cal N} - 1}
\right)}}\langle{\sum\limits_{i \ne j} {\sigma _ + ^i \sigma _ -
^j }}\rangle.
\end{eqnarray}
It is plain that in the absence of the cavity, the correlation
function vanishes and the only   contribution to $I_1$ is given by
the first term,
\begin{equation}
\label{Isl} I_{single}=I_{cl}\equiv I_0\left({1 + e^{\frac
{2\omega}{kT}}}\right)^{-1}.
\end{equation}
The contribution of any remote superzone can easily be found due
to the factorization property of the operator Eq.(\ref{factor}).
If we denote the trace over  states that belong to the same remote
superzone by subscript $K$, we obtain
\begin{eqnarray}
\label{SS} &\langle{S_+ S_-}\rangle_K = \frac {{\cal N}}{2Z} e^
{\frac{K\omega}{kT}} \left({2\cosh\left( {\frac{\Omega_R} {2kT}
}\right) }\right)^{{\cal N}} + \nonumber\\
& \frac {{\cal N}({\cal N}-1)}Z e^ {\frac{K\omega}{kT}}
\left({2\cosh\left( {\frac{\Omega_R} {2kT} }\right)
}\right)^{{\cal N}-2}\left({\sinh\left( {\frac{\Omega_R} {2kT}
}\right) }\right)^2.
\end{eqnarray}
In Fig.~\ref{fig:Int} we compare the spontaneous emission $I_1$ in
the presence of a cavity against the intensity $I_{cl}$ of ${\cal
N}$ atoms in the absence of cavity.

Notice that if the number of atoms is big enough, the intensity of
radiation exhibits a high maximum. In a cavity at low $T$,
 the cluster of ${\cal N}$ two-level atoms emits much
more intensively than it does in the free space. It should be
possible to drive the system to thermal equilibrium at the
temperature where the spontaneous emission exhibits maximum. The
marked amplification of spontaneous emission should be observed in
cavity experiments.

Concluding this section, we recapitulate our main results. We
consider spontaneous emission of the system comprised  of ${\cal
N}$ two-level atoms strongly coupled to the cavity mode and
prepared in the state of thermal equilibrium. In the absence of
the cavity the atoms in the thermal equilibrium would be
uncorrelated. In this  case the spontaneous emission  would be
described by the conventional  formula Eq.(\ref{Isl}). For
  high-Q resonators   the strong coupling to the resonator mode
  should necessarily
be taken into account.  We demonstrate that in this case the
intensity of spontaneous emission can be be greatly enhanced.
 This phenomenon can be explained by additional correlation between
atoms established by the cavity mode. To analyze the effect
analytically we have replaced the exact Tavis-Cummings Hamiltonian
Eq.(\ref{eq1}) by its zero order approximation derived in the
previous sections. This allowed us to represent the intensity of
spontaneous emission in the simple analytical form
Eq.(\ref{I_spon2}). It is appropriate to emphasize once again that
the zero order approximation of the Hamiltonian contains strong
coupling and, thus, describes cooperative effects in the atomic
subsystem. This is the consequence of the fact that the operators
$\tilde S_\pm$ describe collective excitations in the atom-field
system.

\section{Conclusion}
\label{sec:conclusion}

In this work we solved the Tavis-Cummings problem by applying the
technique  of polynomially deformed algebras. We constructed the
transformations that map one polynomial algebra of operators onto
another. This allowed us to reformulate the problem in terms of a
simpler algebra of second order, $\mathbb{S}_r$, and develop a
specific perturbation theory. Our results have a significant
advantage over the so-called linearization approximation, i.e. the
case when the Hamiltonian is linearized in terms of the
 algebra $\mathfrak{U}_1$. In this latter approach a
structure polynomial of a higher order is approximated by a
polynomial of first order. This method allows only to calculate
the lowest or highest eigenvalues, and the corresponding
eigenstates. It does not take into account the finite
dimensionality of the representation. The  parabolic approximation
developed in this work provides in this respect a significant
advantage because it allows to construct a finite dimensional
representation for the problem. We were able to find analytical
expressions for all the eigenvalues of the Hamiltonian up to third
order in the small parameter $\alpha$. For the nearby zones we
showed explicitly how  the collective quantum Rabi frequency
depends on Dicke index $r$. Since this index characterizes the
symmetry of atomic states, the result has significant physical
implications. The dependence on atomic symmetry is revealed
already in zeroth order in the perturbation expansion. Employing
our methods we found an interesting new effect, amplification of
spontaneous emission of thermal ${\cal N}$-atom states due to
collective effects. We expect that this phenomenon can be observed
in cavity experiments. It is worth mentioning that the
applicability of the  method developed in this paper  can be
extended to many other problems, including Bose-Einstein
condensation, multi-photon interactions in the micromaser, and
multimode interaction of an electromagnetic field with matter. Our
methods allow in particular  to address the problem of collective
and dressed states in the mentioned physical systems.

\section*{ACKNOWLEDGEMENTS}

This work has been supported by the Academy of Finland (Project
No. 44875).

I.V. is grateful to the Centre for International Mobility (CIMO)
and University of Jyvaskyla for financial support.

\appendix*

\section{Similarity transformations }
\label{app:SimTran}  We look for similarity transformations
$\tilde {U}_k$ that diagonalize the Hamiltonian in different
orders of $\alpha$,

\begin{equation}
V_{k}^{\left( n \right)}\equiv \tilde {U}_k V_{k-1}^{\left( n
\right)} \tilde {U}_k ^{ - 1}, \label{app0}
\end{equation}

\noindent where $k=0,1,2,3$. In Eq. (\ref{app0}))   only the terms
of order $n$ in the small parameter $\alpha$ are present. For
$k=0$, the term $V_{k-1}^{\left( n \right)}$ should be replaced by
the corresponding term in Eq.(\ref{eq31}). Regrouping the terms we
obtain
\begin{equation}
{\mathop{\rm V}\nolimits} ^{\left( 0 \right)}  = 2\tilde S_x
\left( {1 - \frac{1}{2}\left( {\frac{\alpha }{4}} \right)^2 }
\right), \label{app01}
\end{equation}

\begin{eqnarray}
{\mathop{\rm V}\nolimits} ^{\left( 1 \right)}  =  - \frac{\alpha
}{2}\left( {1 + \frac{1}{4}\left( {\frac{\alpha }{4}} \right)^2 }
\right) \cdot B,\label{app02}
\end{eqnarray}

\begin{equation}
{\mathop{\rm V}\nolimits} ^{\left( 2 \right)}  =  - \left(
{\frac{\alpha }{2}} \right)^2  \cdot \left[ {\tilde S_3
\frac{{\tilde S_ +   + \tilde S_ -  }}{2}\tilde S_3 } \right],
\label{app03}
\end{equation}

\begin{equation}
{\mathop{\rm V}\nolimits} ^{\left( 3 \right)}  =  -
\frac{1}{2}\left( {\frac{\alpha }{2}} \right)^3 \left[ {\tilde S_3
\cdot B \cdot \tilde S_3 } \right], \label{app04}
\end{equation}
where
\begin{equation}
B \equiv \left[ {\tilde S_3 \tilde S_x + \tilde S_x \tilde S_3 }
\right], \label{app001}
\end{equation}
and
\begin{equation}
\tilde S_x\equiv \frac{{\tilde S_ +   + \tilde S_ - }}{2}.
\label{app002}
\end{equation}

\subsection{\label{app:subsec1} The zero order transformation $\tilde {U}_0$}
As known from the theory of \textit{su(2)} algebra, the operator
$\tilde S_x$  can be diagonalized by the transformation
\begin{equation}
\tilde {U}_0 = \exp \left[ {\frac{\pi }{4}\left( {\tilde{S}_ + -
\tilde{S}_ - } \right)} \right]= \exp \left[ -i{\frac{\pi
}{2}\tilde S_y} \right]. \label{app1}
\end{equation}

\noindent Employing this transformation we obtain

\begin{eqnarray}
{\mathop{\rm V}\nolimits} _0^{\left( 0 \right)}  = 2\tilde S_3
  \left( {1 - \frac{1}{2}\left( {\frac{\alpha }{4}} \right)^2 } \right),
  \;{\mathop{\rm V}\nolimits} _0^{\left( 1 \right)}  =
   - {\mathop{\rm V}\nolimits} ^{\left( 1 \right)} ,\;\nonumber\\
   {\mathop{\rm V}\nolimits} _0^{\left( 2 \right)}  =
    - \left( {\frac{\alpha }{2}} \right)^2  \cdot
\left[ {\tilde S_x \tilde S_3 \tilde S_x} \right],\nonumber\\
    {\mathop{\rm V}\nolimits} _0^{\left( 3 \right)}  = \frac{1}{2}
  \left( {\frac{\alpha }{2}} \right)^3 \left[ {\tilde S_x
  B \tilde S_x } \right].
\label{app2} \end{eqnarray}

\subsection{\label{app:subsec2} The first order transformation $\tilde {U}_1$}

It  can readily be seen that the transformation
\begin{equation}
\tilde U_1  = \exp \left[ {\alpha x \cdot D1} \right],\quad D1
\equiv -i\left[ {\tilde S_3 \tilde S_y + \tilde S_y\tilde S_3 }
\right] \label{app6}
\end{equation}
\noindent diagonalizes the operators in the first order. In the
diagonalization one needs the commutators

\begin{widetext}
\begin{eqnarray}
\left[ {\tilde S_3 ,D1} \right] = B, \nonumber\\
  \left[ {\left[ {\tilde S_3 ,D1} \right],D1} \right] =
  4\tilde S_3 \left( {\tilde S^2  - 2\tilde S_3 ^2  - \frac{1}{4}}
  \right),\nonumber\\
  \left[ {\left[ {\left[ {\tilde S_3 ,D1} \right],D1} \right],D1} \right] =
  \left( {4\tilde S^2  - 1} \right) B - 8
  \left[ {\tilde S_3 B + B\tilde S_3  - \tilde S_3
  \tilde S_x \tilde S_3 }
  \right],\nonumber\\
  \left[ {\tilde S_3 \tilde S_x \tilde S_3 ,D1} \right]
   = 2\tilde S_3 ^2 \left( {\tilde S^2  - 2\tilde S_3 ^2 } \right) +
    \left( {\tilde S_3 \tilde S_x} \right)^2  +
    \left( {\tilde S_x\tilde S_3 } \right)^2.
\label{app6_2}
\end{eqnarray}
\end{widetext}

Then up to third order in $\alpha$,

\begin{eqnarray}
\tilde U_1 {\mathop{\rm V}\nolimits} _0^{\left( 0 \right)} \left(
{\tilde U_1 } \right)^{ - 1}  = {\mathop{\rm V}\nolimits}
_0^{\left( 0 \right)}  - 2\alpha x\left( {1 - \frac{1}{2}\left(
{\frac{\alpha }{4}} \right)^2 } \right)B + \nonumber\\
4\left( {\alpha x} \right)^2 \tilde S_3 \left( {\tilde S^2  -
2\tilde S_3 ^2  - \frac{1}{4}} \right) -
\frac{{\left( {\alpha x} \right)^3 }}{3}\cdot \nonumber\\
\left\{ {\left( {4\tilde S^2  - 1} \right) B - 8 \tilde J}
\right\}, \label{app8}
\end{eqnarray}

\noindent and

\begin{eqnarray}
&\tilde U_1 {\mathop{\rm V}\nolimits} _0^{\left( 1 \right)} \left(
{\tilde U_1 } \right)^{ - 1}  = \frac{\alpha }{2}\left( {1 +
\frac{1}{4}\left( {\frac{\alpha }{4}} \right)^2 } \right) \cdot B
- \nonumber\\
&\alpha^2 x \cdot 2\tilde S_3 \left( {\tilde S^2 - 2\tilde S_3 ^2
- \frac{1}{4}} \right) +  \alpha^3\left( {\frac{x }{2}} \right)^2
 \cdot \nonumber\\
&\left\{ {\left( {4\tilde S^2 - 1} \right)B - 8 \tilde J}
\right\},\label{app9}
\end{eqnarray}

\noindent where $\tilde J\equiv \tilde S_3 B + B\tilde S_3 -
\tilde S_3 \tilde S_x\tilde S_3 $.

\begin{widetext}
\begin{eqnarray}
&\tilde U_1 {\mathop{\rm V}\nolimits} _0^{\left( 2 \right)} \left(
{\tilde U_1 } \right)^{ - 1}  = {\mathop{\rm V}\nolimits}
_0^{\left( 2 \right)}  +  \nonumber\\
&\alpha^3\left( {\frac{x }{2}} \right)^2 \cdot \left\{ {\tilde S_x
\tilde S_3 \left( {2\tilde S^2  - 4\tilde S_3 ^2  - \tilde S_x^2 }
\right) + \left( {2\tilde S^2  - 4\tilde S_3 ^2  - \tilde S_x^2 }
\right)\tilde S_3 \tilde S_x} \right\} \label{app10}
\end{eqnarray}
\end{widetext}

\noindent In third order the operator $V_0^{\left( 3 \right)} $
remains unchanged after the transformation, i.e. $V_1^{\left( 3
\right)} = V_0^{\left( 3 \right)} $. In order to calculate
$V_0^{\left( 2 \right)} $ up to third order, we take into account
that

$$x = \frac{1}{4} \frac {\left( {1 + \left( {\frac{\alpha }{8}}
\right)^2 } \right)} {\left( {1 - 2\left( {\frac{\alpha }{8}}
\right)^2 } \right)} \approx \frac{1}{4} \cdot \left( {1 + 3\left(
{\frac{\alpha }{8}} \right)^2 } \right),$$

and find then that

\begin{widetext}
\begin{eqnarray}
&{\mathop{\rm V}\nolimits} _1^{\left( 0 \right)}  = {\mathop{\rm
V}\nolimits} _0^{\left( 0 \right)}  - 4\left( {\frac{\alpha }{4}}
\right)^2 \tilde S_3 \left( {\tilde S^2  - 2\tilde S_3 ^2  -
\frac{1}{4}} \right),\quad {\mathop{\rm V}\nolimits} _1^{\left( 1
\right)}  = 0,\quad {\mathop{\rm V}\nolimits} _1^{\left( 2
\right)}  = {\mathop{\rm V}\nolimits} _0^{\left( 2 \right)} ,
\nonumber\\
&{\mathop{\rm V}\nolimits} _1^{\left( 3 \right)}  = {\mathop{\rm
V}\nolimits} _0^{\left( 3 \right)}  - \frac{1}{2}\left(
{\frac{\alpha }{4}} \right)^3 B + \frac{2}{3}\left( {\frac{\alpha
}{4}} \right)^3 \left\{ {\left( {4\tilde S^2  - 1} \right) \cdot B
- 8 \tilde J}\right\} +  \nonumber\\
&\left( {\frac{\alpha }{4}} \right)^3 \cdot \left\{ {\tilde
S_x\tilde S_3 \left( {2\tilde S^2  - 4\tilde S_3 ^2  - \left(
{\tilde S_x} \right)^2 } \right) + \left( {2\tilde S^2  - 4\tilde
S_3 ^2 - \left( {\tilde S_x} \right)^2 } \right)\tilde S_3 \tilde
S_x} \right\}. \label{app11}
\end{eqnarray}
\end{widetext}

\subsection{\label{app:subsec3} The second and the third orders transformations}

To find the  second order transformation we rewrite ${\mathop{\rm
V}\nolimits} _1^{\left( 2 \right)}$ in a symmetrized form,

\begin{equation}
{\mathop{\rm V}\nolimits}_1^{\left( 2 \right)}  =  - \frac12
\left( {\frac{\alpha }{2}} \right)^2  \left[ {\tilde L_x\tilde S_3
+ \tilde S_3 \tilde L_x + \tilde S_3 \left( {\tilde S^2  - \tilde
S_3 ^2  - 1} \right)} \right], \label{app12}
\end{equation}

\noindent where $\tilde L_x \equiv \frac{{\tilde S_ + ^2 + \tilde
S_ -  ^2 }}{4}$. The diagonalizing transformation is then given by
\begin{equation}
\tilde U_2  \equiv \exp \left\{ { i\frac12\left( {\frac{\alpha
}{4}} \right)^2  \cdot \left[ {\tilde L_y\tilde S_3  + \tilde S_3
\tilde L_y} \right]} \right\}, \label{app13}
\end{equation}
\noindent where $\tilde L_y \equiv \frac{{\tilde S_ + ^2 - \tilde
S_ -  ^2 }}{4}$. Keeping the terms up to third order we obtain
(see (\ref{app9}))
\begin{eqnarray}
\tilde U_2 {\mathop{\rm V}\nolimits} _1^{\left( 0 \right)} \left(
{\tilde U_2 } \right)^{ - 1}  = {\mathop{\rm V}\nolimits}
_1^{\left( 0 \right)}  + \frac12\left( {\frac{\alpha }{2}}
\right)^2 \left[ {\tilde L_x\tilde S_3 + \tilde S_3 \tilde L_x}
\right].  \label{app14}
\end{eqnarray}

\noindent The transformation (\ref{app13})  does not change the
expressions, given above for$V_1^{\left( 2 \right)} $ and
$V_1^{\left( 3 \right)} $, and we find that

\begin{eqnarray}
&{\mathop{\rm V}\nolimits} _2^{\left( 0 \right)}  = {\mathop{\rm
V}\nolimits} _1^{\left( 0 \right)}  - \left( {\frac{\alpha }{4}}
\right)^2  \cdot 2\tilde S_3 \left( {\tilde S^2  - \tilde S_3 ^2 -
1} \right),\nonumber\\
&{\mathop{\rm V}\nolimits} _2^{\left( 1 \right)} = {\mathop{\rm
V}\nolimits} _2^{\left( 2 \right)}  = 0,\quad {\mathop{\rm
V}\nolimits} _2^{\left( 3 \right)}  = {\mathop{\rm V}\nolimits}
_1^{\left( 3 \right)} .
 \label{app15}
\end{eqnarray}

Diagonalization of $V_2^{\left( 3 \right)}$ can be performed in a
similar way with an operator $\tilde U_3 = \exp \left[ { - \left(
{\frac{\alpha }{4}} \right)^3 \cdot \mathcal{O}} \right]$. As
there are no diagonal terms in $V_2^{\left( 3 \right)}$, which
would contribute to the spectrum of the Hamiltonian, we do not
give here the fairly complicated form of operator $\mathcal{O}$.
The final diagonal form for the interaction $V$ is thus given by

\begin{eqnarray}
\bar V  = \Omega  \tilde S_3 \left\{ {1 + \left( {\frac{\alpha
}{4}} \right)^2  \left[ {5\tilde S_3 ^2  - 3\tilde r\left( {\tilde
r + 1} \right) + 1} \right]} \right\}.
 \label{app17}
\end{eqnarray}

To recapitulate, we introduced four transformations $\tilde {U}_k
,\quad k = 0,1,2,3$, which successively diagonalize the
interaction operator in the Tavis-Cummings Hamiltonian up to third
order with respect to the small parameter
$\alpha=\left({\frac12\Omega_R}\right)^{-2}$, with $\Omega_R$ the
generalized Rabi frequency of Eq.(\ref{eq_Rabi}), such that

\begin{eqnarray}
\bar V \equiv \tilde {U} V  \tilde {U}^{ - 1},\quad \tilde {U}
\equiv \tilde {U}_3  \tilde {U}_2  \tilde {U}_1  \tilde {U}_0.
\label{app18}
\end{eqnarray}


\newpage

\begin{thebibliography}{15}
\expandafter\ifx\csname
natexlab\endcsname\relax\def\natexlab#1{#1}\fi
\expandafter\ifx\csname bibnamefont\endcsname\relax
  \def\bibnamefont#1{#1}\fi
\expandafter\ifx\csname bibfnamefont\endcsname\relax
  \def\bibfnamefont#1{#1}\fi
\expandafter\ifx\csname citenamefont\endcsname\relax
  \def\citenamefont#1{#1}\fi
\expandafter\ifx\csname url\endcsname\relax
  \def\url#1{\texttt{#1}}\fi
\expandafter\ifx\csname
urlprefix\endcsname\relax\def\urlprefix{URL }\fi
\providecommand{\bibinfo}[2]{#2}
\providecommand{\eprint}[2][]{\url{#2}}

\bibitem[{\citenamefont{Tavis and Cummings}(1968)}]{Tavis:1968}
\bibinfo{author}{\bibfnamefont{M.}~\bibnamefont{Tavis}} \bibnamefont{and}
  \bibinfo{author}{\bibfnamefont{F.~W.} \bibnamefont{Cummings}},
  \bibinfo{journal}{Phys.\ Rev.} \textbf{\bibinfo{volume}{170}},
  \bibinfo{pages}{379} (\bibinfo{year}{1968}).

\bibitem[{\citenamefont{Dicke}(1954)}]{Dicke:1954}
\bibinfo{author}{\bibfnamefont{R.}~\bibnamefont{Dicke}},
  \bibinfo{journal}{Phys.\ Rev.} \textbf{\bibinfo{volume}{93}},
  \bibinfo{pages}{99} (\bibinfo{year}{1954}).

\bibitem[{\citenamefont{Jaynes and Cummings}(1963)}]{Jaynes:1963}
\bibinfo{author}{\bibfnamefont{E.}~\bibnamefont{Jaynes}} \bibnamefont{and}
  \bibinfo{author}{\bibfnamefont{F.}~\bibnamefont{Cummings}},
  \bibinfo{journal}{Proc. IEEE} \textbf{\bibinfo{volume}{51}},
  \bibinfo{pages}{89} (\bibinfo{year}{1963}).

\bibitem[{\citenamefont{Meschede et~al.}(1985)\citenamefont{Meschede, Walther,
  and Muller}}]{Meschede:1985}
\bibinfo{author}{\bibfnamefont{D.}~\bibnamefont{Meschede}},
  \bibinfo{author}{\bibfnamefont{H.}~\bibnamefont{Walther}}, \bibnamefont{and}
  \bibinfo{author}{\bibfnamefont{G.}~\bibnamefont{Muller}},
  \bibinfo{journal}{Phys. Rev. Lett.} \textbf{\bibinfo{volume}{54}},
  \bibinfo{pages}{551} (\bibinfo{year}{1985}).

\bibitem[{\citenamefont{Filipowicz et~al.}(1986)\citenamefont{Filipowicz,
  Javanainen, and Meystre}}]{Filipowicz:1996}
\bibinfo{author}{\bibfnamefont{P.}~\bibnamefont{Filipowicz}},
  \bibinfo{author}{\bibfnamefont{J.}~\bibnamefont{Javanainen}},
  \bibnamefont{and} \bibinfo{author}{\bibfnamefont{P.}~\bibnamefont{Meystre}},
  \bibinfo{journal}{Phys. Rev. A} \textbf{\bibinfo{volume}{34}},
  \bibinfo{pages}{3077} (\bibinfo{year}{1986}).

\bibitem[{\citenamefont{Elmfors et~al.}(1996)\citenamefont{Elmfors, Lautrup,
  and Skagerstam}}]{Elmfors:1996}
\bibinfo{author}{\bibfnamefont{P.}~\bibnamefont{Elmfors}},
  \bibinfo{author}{\bibfnamefont{B.}~\bibnamefont{Lautrup}}, \bibnamefont{and}
  \bibinfo{author}{\bibfnamefont{B.-S.} \bibnamefont{Skagerstam}},
  \bibinfo{journal}{Phys. Rev. A} \textbf{\bibinfo{volume}{54}},
  \bibinfo{pages}{5171} (\bibinfo{year}{1996}).

\bibitem[{\citenamefont{Raimond}(2001)}]{Raimond:2001}
\bibinfo{author}{\bibfnamefont{J.M.}~\bibnamefont{Raimond}},
  \bibinfo{author}{\bibfnamefont{M.}~\bibnamefont{Brune}}, \bibnamefont{and}
  \bibinfo{author}{\bibfnamefont{S.} \bibnamefont{Haroche}},
  \bibinfo{journal}{Rev. Mod. Phys.} \textbf{\bibinfo{volume}{73}},
  \bibinfo{pages}{565} (\bibinfo{year}{2001}).

\bibitem[{\citenamefont{Yu et~al.}(1995)\citenamefont{Yu, Rauch, and
  Zhang}}]{Sixia:1997}
\bibinfo{author}{\bibfnamefont{S.}~\bibnamefont{Yu}},
  \bibinfo{author}{\bibfnamefont{H.}~\bibnamefont{Rauch}}, \bibnamefont{and}
  \bibinfo{author}{\bibfnamefont{Y.}~\bibnamefont{Zhang}},
  \bibinfo{journal}{Phys. Rev. A} \textbf{\bibinfo{volume}{52}},
  \bibinfo{pages}{2585} (\bibinfo{year}{1995}).

\bibitem[{\citenamefont{Yang et~al.}(1997)\citenamefont{Yang, Wu, and
  Li}}]{Yang}
\bibinfo{author}{\bibfnamefont{X.}~\bibnamefont{Yang}},
  \bibinfo{author}{\bibfnamefont{Y.}~\bibnamefont{Wu}}, \bibnamefont{and}
  \bibinfo{author}{\bibfnamefont{Y.}~\bibnamefont{Li}}, \bibinfo{journal}{Phys.
  Rev. A} \textbf{\bibinfo{volume}{55}}, \bibinfo{pages}{4545}
  (\bibinfo{year}{1997}).

\bibitem[{\citenamefont{Rybin}(1998)\citenamefont{Rybin}}]{Rybin:1998}
\bibinfo{author}{\bibfnamefont{A.}~\bibnamefont{Rybin}},
  \bibinfo{author}{\bibfnamefont{G.}~\bibnamefont{Kastelewicz}},
  \bibinfo{author}{\bibfnamefont{J.}~\bibnamefont{Timonen}}, \bibnamefont{and}
  \bibinfo{author}{\bibfnamefont{N.}~\bibnamefont{Bogoliubov}},
  \bibinfo{journal}{J. Phys. A} \textbf{\bibinfo{volume}{31}}, \bibinfo{pages}{4705}
  (\bibinfo{year}{1998});
  \bibinfo{author}{\bibfnamefont{N.M.}~\bibnamefont{Bogoliubov}},
    \bibinfo{author}{\bibfnamefont{R.K.}~\bibnamefont{Bullough}}, \bibnamefont{and}
  \bibinfo{author}{\bibfnamefont{J.}~\bibnamefont{Timonen}},
  \bibinfo{journal}{J. Phys. A} \textbf{\bibinfo{volume}{29}}, \bibinfo{pages}{6305}
  (\bibinfo{year}{1996});

\bibitem[{\citenamefont{Carusotto}(1989)}]{Carusotto:1989}
\bibinfo{author}{\bibfnamefont{S.}~\bibnamefont{Carusotto}}, \bibinfo{journal}{
  Phys. Rev. A} \textbf{\bibinfo{volume}{40}}, \bibinfo{pages}{1848}
  (\bibinfo{year}{1989}).

\bibitem[{\citenamefont{Smithers}(1974)}]{Smithers:1974}
\bibinfo{author}{\bibfnamefont{M.E.}~\bibnamefont{Smithers}}, \bibnamefont{and}
  \bibinfo{author}{\bibfnamefont{E.Y.C.}~\bibnamefont{Lu}}, \bibinfo{journal}{
  Phys. Rev. A} \textbf{\bibinfo{volume}{9}}, \bibinfo{pages}{790}
  (\bibinfo{year}{1974}).

\bibitem[{\citenamefont{Mollow}(1967)}]{Mollow:1967}
\bibinfo{author}{\bibfnamefont{B.R.}~\bibnamefont{Mollow}}, \bibnamefont{and}
  \bibinfo{author}{\bibfnamefont{R.J.}~\bibnamefont{Glauber}}, \bibinfo{journal}{
  Phys. Rev.} \textbf{\bibinfo{volume}{160}}, \bibinfo{pages}{1076}
  (\bibinfo{year}{1967}).

\bibitem[{\citenamefont{Tucker}(1967)}]{Tucker:1969}
\bibinfo{author}{\bibfnamefont{J.}~\bibnamefont{Tucker}}, \bibnamefont{and}
  \bibinfo{author}{\bibfnamefont{D.F.}~\bibnamefont{Walls}}, \bibinfo{journal}{
  Phys. Rev.} \textbf{\bibinfo{volume}{178}}, \bibinfo{pages}{2036}
  (\bibinfo{year}{1969}).

\bibitem[{\citenamefont{Klimov}(2000)}]{Klimov:2000}
\bibinfo{author}{\bibfnamefont{A.B.}~\bibnamefont{Klimov}}, \bibnamefont{and}
  \bibinfo{author}{\bibfnamefont{L.L.}~\bibnamefont{Sanchez-Soto}}, \bibinfo{journal}{
  Phys. Rev. A} \textbf{\bibinfo{volume}{61}}, \bibinfo{pages}{063802}
  (\bibinfo{year}{2000}).

\bibitem[{\citenamefont{Kozierowski}(1990)}]{Kozierowski:1990}
\bibinfo{author}{\bibfnamefont{M.}~\bibnamefont{Kozierowski}}, \bibnamefont{and}
  \bibinfo{author}{\bibfnamefont{A.A.}~\bibnamefont{Mamedov}}, \bibnamefont{and}
  \bibinfo{author}{\bibfnamefont{S.M.}~\bibnamefont{Chumakov}}, \bibinfo{journal}{
  Phys. Rev. A} \textbf{\bibinfo{volume}{42}}, \bibinfo{pages}{1762}
  (\bibinfo{year}{1990}).

\bibitem[{\citenamefont{Chumakov}(1994)}]{Chumakov:1994}
\bibinfo{author}{\bibfnamefont{S.M.}~\bibnamefont{Chumakov}}, \bibnamefont{and}
  \bibinfo{author}{\bibfnamefont{A.B.}~\bibnamefont{Klimov}}, \bibnamefont{and}
  \bibinfo{author}{\bibfnamefont{J.J.}~\bibnamefont{Sanchez-Mondragon}}, \bibinfo{journal}{
  Phys. Rev. A} \textbf{\bibinfo{volume}{49}}, \bibinfo{pages}{4972}
  (\bibinfo{year}{1994}).

\bibitem[{\citenamefont{Saavedra}(1998)}]{Saavedra:1998}
  \bibinfo{author}{\bibfnamefont{C.}~\bibnamefont{Saavedra}}, \bibnamefont{and}
  \bibinfo{author}{\bibfnamefont{A.B.}~\bibnamefont{Klimov}}, \bibnamefont{and}
  \bibinfo{author}{\bibfnamefont{S.M.}~\bibnamefont{Chumakov}}, \bibnamefont{and}
  \bibinfo{author}{\bibfnamefont{J.C.}~\bibnamefont{Retamal}}, \bibinfo{journal}{
  Phys. Rev. A} \textbf{\bibinfo{volume}{58}}, \bibinfo{pages}{4078}
  (\bibinfo{year}{1998}).

  \bibitem[{\citenamefont{Delgado}(2001)}]{Delgado:2001}
  \bibinfo{author}{\bibfnamefont{J.}~\bibnamefont{Delgado}}, \bibnamefont{and}
  \bibinfo{author}{\bibfnamefont{E.C.}~\bibnamefont{Yustas}}, \bibnamefont{and}
  \bibinfo{author}{\bibfnamefont{L.L.}~\bibnamefont{Sanchez-Soto}}, \bibinfo{journal}{
  \bibinfo{author}{\bibfnamefont{A.B.}~\bibnamefont{Klimov}}, \bibnamefont{and}
  Phys. Rev. A} \textbf{\bibinfo{volume}{63}}, \bibinfo{pages}{063801}
  (\bibinfo{year}{2001}).



\bibitem[{\citenamefont{Brief}(1996)}]{Brief:1996}
\bibinfo{author}{\bibfnamefont{C.}~\bibnamefont{Brif}}, \bibinfo{journal}{
  Phys. Rev. A} \textbf{\bibinfo{volume}{54}}, \bibinfo{pages}{5253}
  (\bibinfo{year}{1996}).

\bibitem[{\citenamefont{Sunilkumar}(2000)}]{Sunilkumar:2000}
\bibinfo{author}{\bibfnamefont{V.}~\bibnamefont{Sunilkumar}},
\bibinfo{author}{\bibfnamefont{B.A.}~\bibnamefont{Bambah}},
\bibinfo{author}{\bibfnamefont{R.}~\bibnamefont{Jagannathan}},
\bibinfo{author}{\bibfnamefont{P.K.}~\bibnamefont{Panigrahi}},
\bibnamefont{and}\bibinfo{author}{\bibfnamefont{V.}~\bibnamefont{Srinivasan}},
\bibinfo{journal}{J.Opt.B } \textbf{\bibinfo{volume}{2}}, \bibinfo{pages}{126}
  (\bibinfo{year}{2000}).

\bibitem[{\citenamefont{Holstein and Primakoff}(1940)}]{Holstein:1940}
\bibinfo{author}{\bibfnamefont{T.}~\bibnamefont{Holstein}} \bibnamefont{and}
  \bibinfo{author}{\bibfnamefont{H.}~\bibnamefont{Primakoff}},
  \bibinfo{journal}{Phys. Rev.} \textbf{\bibinfo{volume}{58}},
  \bibinfo{pages}{1098} (\bibinfo{year}{1940}).

\bibitem[{\citenamefont{Persico}(1975)}]{Persico:1975}
\bibinfo{author}{\bibfnamefont{F.}~\bibnamefont{Persico}}, \bibnamefont{and}
  \bibinfo{author}{\bibfnamefont{G.}~\bibnamefont{Vetri}},
  \bibinfo{journal}{Phys. Rev. A} \textbf{\bibinfo{volume}{12}},
  \bibinfo{pages}{2083} (\bibinfo{year}{1975}).

\bibitem[{\citenamefont{Daskaloyannis}(1993)\citenamefont{Bonatsos, Daskaloyannis, and
  Lalazissis}}]{Daskaloyannis:1993}
\bibinfo{author}{\bibfnamefont{D.}~\bibnamefont{Bonatsos}},
  \bibinfo{author}{\bibfnamefont{C.}~\bibnamefont{Daskaloyannis}},
  \bibinfo{author}{\bibfnamefont{P.}~\bibnamefont{Kolokotronis}},
  \bibinfo{journal}{J. Phys. A} \textbf{\bibinfo{volume}{26}},
  \bibinfo{pages}{L871} (\bibinfo{year}{1993}).

\bibitem[{\citenamefont{Ruan}()}]{Ruan:2001}
\bibinfo{author}{\bibfnamefont{D.}~\bibnamefont{Ruan}},
  \bibinfo{author}{\bibfnamefont{Y.}~\bibnamefont{Jia}}, \bibnamefont{and}
  \bibinfo{author}{\bibfnamefont{H.}~\bibnamefont{Sun}},
  \eprint{quant-ph/0111007}.


\bibitem[{\citenamefont{Karassiov}(1994)}]{Karassiov:1994}
\bibinfo{author}{\bibfnamefont{V.P.}~\bibnamefont{Karassiov}},
  \bibinfo{journal}{J. Phys. A} \textbf{\bibinfo{volume}{27}},
  \bibinfo{pages}{153} (\bibinfo{year}{1994});
  \bibinfo{author}{\bibfnamefont{V.P.}~\bibnamefont{Karassiov}},
  \bibinfo{author}{\bibfnamefont{A.A.}~\bibnamefont{Gusev}}, \bibnamefont{and}
  \bibinfo{author}{\bibfnamefont{S.I.}~\bibnamefont{Vinitsky}},
  \bibinfo{preprint}{quant-ph/0112040}(\bibinfo{year}{2001}).


\bibitem[{\citenamefont{Bonatsos et~al.}(1993)\citenamefont{Bonatsos, Daskaloyannis, and
  Lalazissis}}]{Bonatsos:1993}
\bibinfo{author}{\bibfnamefont{D.}~\bibnamefont{Bonatsos}},
  \bibinfo{author}{\bibfnamefont{C.}~\bibnamefont{Daskaloyannis}},
  \bibinfo{author}{\bibfnamefont{G.~A.}~\bibnamefont{Lalazissis}},
  \bibinfo{journal}{Phys. Rev. A} \textbf{\bibinfo{volume}{47}},
  \bibinfo{pages}{3448} (\bibinfo{year}{1993}).


\bibitem[{\citenamefont{Brandt and Greenberg}(1969)}]{Brandt:1979}
\bibinfo{author}{\bibfnamefont{R.}~\bibnamefont{Brandt}} \bibnamefont{and}
  \bibinfo{author}{\bibfnamefont{O.~W.} \bibnamefont{Greenberg}},
  \bibinfo{journal}{J. Math. Phys.} \textbf{\bibinfo{volume}{10}},
  \bibinfo{pages}{1168} (\bibinfo{year}{1969}).

\bibitem[{\citenamefont{Rasetti}(1972)}]{Rasetti:1972}
\bibinfo{author}{\bibfnamefont{M.}~\bibnamefont{Rasetti}},
  \bibinfo{journal}{Int. J. Theor. Phys.} \textbf{\bibinfo{volume}{5}},
  \bibinfo{pages}{377} (\bibinfo{year}{1972}).

\bibitem[{\citenamefont{Katriel}(1979)}]{Katriel:1979}
\bibinfo{author}{\bibfnamefont{J.}~\bibnamefont{Katriel}},
  \bibinfo{journal}{Phys. Lett.} \textbf{\bibinfo{volume}{72A}},
  \bibinfo{pages}{94} (\bibinfo{year}{1979}).

\bibitem[{\citenamefont{Gerry}(1971)}]{Gerry:1971}
\bibinfo{author}{\bibfnamefont{C.}~\bibnamefont{Gerry}}, \bibinfo{journal}{J.
  Phys. A} \textbf{\bibinfo{volume}{16}}, \bibinfo{pages}{313}
  (\bibinfo{year}{1971}).

\bibitem[{\citenamefont{Katriel et~al.}(1986)\citenamefont{Katriel, Solomon,
  DAriano, and Rasetti}}]{Katriel:1986}
\bibinfo{author}{\bibfnamefont{J.}~\bibnamefont{Katriel}},
  \bibinfo{author}{\bibfnamefont{A.~I.}~\bibnamefont{Solomon}},
  \bibinfo{author}{\bibfnamefont{G.}~\bibnamefont{DAriano}}, \bibnamefont{and}
  \bibinfo{author}{\bibfnamefont{M.}~\bibnamefont{Rasetti}},
  \bibinfo{journal}{Phys. Rev. D} \textbf{\bibinfo{volume}{34}},
  \bibinfo{pages}{2332} (\bibinfo{year}{1986}).

\bibitem[{\citenamefont{Carusotto}(1988)}]{Carusotto:1988}
\bibinfo{author}{\bibfnamefont{S.}~\bibnamefont{Carusotto}}, \bibinfo{journal}
{Phys. Rev. A} \textbf{\bibinfo{volume}{38}},
\bibinfo{pages}{3249}(\bibinfo{year}{1971}).

\bibitem[{\citenamefont{Lenis}(1993)}]{Lenis:1993}
\bibinfo{author}{\bibfnamefont{D.}~\bibnamefont{Bonatsos}},
  \bibinfo{author}{\bibfnamefont{C.}~\bibnamefont{Daskaloyannis}},
  \bibinfo{author}{\bibfnamefont{P.}~\bibnamefont{Kolokotronis}}, \bibnamefont{and}
    \bibinfo{author}{\bibfnamefont{D.}~\bibnamefont{Lenis}},
  \eprint{hcp-th/9402099}.

\bibitem[{\citenamefont{Shanta}(1994)}]{Shanta:1994}
\bibinfo{author}{\bibfnamefont{P.}~\bibnamefont{Shanta}},
\bibinfo{author}{\bibfnamefont{S.}~\bibnamefont{Chaturvedi}},
\bibinfo{author}{\bibfnamefont{V.}~\bibnamefont{Srinivasan}},
    \bibinfo{author}{\bibfnamefont{G.S.}~\bibnamefont{Agarwal}}, \bibnamefont{and}
    \bibinfo{author}{\bibfnamefont{C.L.}~\bibnamefont{Mehta}},
    \bibinfo{journal}{Phys. Rev. Lett.} \textbf{\bibinfo{volume}{72}},
\bibinfo{pages}{1447}(\bibinfo{year}{1994}).

\bibitem[{\citenamefont{Floreanini}(1996)}]{Floreanini:1996}
\bibinfo{author}{\bibfnamefont{R.}~\bibnamefont{Floreanini}},
\bibinfo{author}{\bibfnamefont{L.}~\bibnamefont{Lapointe}},
\bibinfo{author}{\bibfnamefont{L.}~\bibnamefont{Vinet}},
    \bibinfo{journal}{Phys. Lett. B} \textbf{\bibinfo{volume}{389}},
\bibinfo{pages}{327}(\bibinfo{year}{1996}).

\bibitem[{\citenamefont{Ruan}(1999)}]{Ruan:1999}
\bibinfo{author}{\bibfnamefont{D.}~\bibnamefont{Ruan}},
\bibinfo{author}{\bibfnamefont{W.}~\bibnamefont{Ruan}},
    \bibinfo{journal}{Phys. Lett. A} \textbf{\bibinfo{volume}{263}},
\bibinfo{pages}{78}(\bibinfo{year}{1999}).


\bibitem[{\citenamefont{Glauber}(1963)}]{Glauber:1963}
\bibinfo{author}{\bibfnamefont{R.J.}~\bibnamefont{Glauber}},
    \bibinfo{journal}{Phys. Rev.} \textbf{\bibinfo{volume}{131}},
\bibinfo{pages}{2766}(\bibinfo{year}{1963}).

\bibitem[{\citenamefont{Rybin}(1999)}]{Rybin:1999}
\bibinfo{author}{\bibfnamefont{A.}~\bibnamefont{Rybin}},
  \bibinfo{author}{\bibfnamefont{G.P.}~\bibnamefont{Miroshnichenko}},
  \bibinfo{author}{\bibfnamefont{I.P.}~\bibnamefont{Vadeiko}}, \bibnamefont{and}
  \bibinfo{author}{\bibfnamefont{J.}~\bibnamefont{Timonen}},
  \bibinfo{journal}{J. Phys. A} \textbf{\bibinfo{volume}{32}},
  \bibinfo{pages}{8739} (\bibinfo{year}{1999}).

\bibitem[{\citenamefont{Witschel}(1974)}]{Witschel:1974}
\bibinfo{author}{\bibfnamefont{W.}~\bibnamefont{Witschel}},
  \bibinfo{journal}{J. Phys. A} \textbf{\bibinfo{volume}{7}},
  \bibinfo{pages}{1847} (\bibinfo{year}{1974}).

  \bibitem[{\citenamefont{Scully}(1998)}]{Scully:1998}
\bibinfo{author}{\bibfnamefont{M.O.}~\bibnamefont{Scully}}
\bibnamefont{and}
\bibinfo{author}{\bibfnamefont{M.S.}~\bibnamefont{Zubairy}},
\bibinfo{title}{Quantum Optics},
  \bibinfo{book}{Cambridge University Press},
\bibinfo{year}{1997}.
\end{thebibliography}
\end{document}